\documentclass[showpacs,aps,prd,nofootinbib,floatfix,amsmath,amssymb]{revtex4}
\usepackage{graphicx}

\begin{document}

\makeatletter
\newbox\slashbox \setbox\slashbox=\hbox{$/$}
\newbox\Slashbox \setbox\Slashbox=\hbox{\large$/$}
\def\pFMslash#1{\setbox\@tempboxa=\hbox{$#1$}
  \@tempdima=0.5\wd\slashbox \advance\@tempdima 0.5\wd\@tempboxa
  \copy\slashbox \kern-\@tempdima \box\@tempboxa}
\def\pFMSlash#1{\setbox\@tempboxa=\hbox{$#1$}
  \@tempdima=0.5\wd\Slashbox \advance\@tempdima 0.5\wd\@tempboxa
  \copy\Slashbox \kern-\@tempdima \box\@tempboxa}
\def\FMslash{\protect\pFMslash}
\def\FMSlash{\protect\pFMSlash}
\def\miss#1{\ifmmode{/\mkern-11mu #1}\else{${/\mkern-11mu #1}$}\fi}
\makeatother

\title{ One-loop effects of extra dimensions on the $WW\gamma$ and $WWZ$ vertices}
\author{A. Flores-Tlalpa$^{(a)}$, J. Monta\~no$^{(b)}$, H. Novales-S\' anchez$^{(a)}$, F. Ram\'\i rez-Zavaleta$^{(c)}$, J. J. Toscano$^{(a)}$}
\address{$^{(a)}$Facultad de Ciencias F\'{\i}sico Matem\'aticas,
Benem\'erita Universidad Aut\'onoma de Puebla, Apartado Postal
1152, Puebla, Puebla, M\'exico. \\
$^{(b)}$Departamento de F\'\i sica, CINVESTAV, A. P. 14-740, 07000, M\' exico, D. F., M\' exico.\\
$^{(c)}$Facultad de Ciencias F\'\i sico Matem\' aticas,
Universidad Michoacana de San Nicol\' as de
Hidalgo, Avenida Francisco J. M\' ujica S/N, 58060, Morelia, Michoac\'an, M\' exico.}
\begin{abstract}
The one-loop contribution of the excited Kaluza-Klein (KK) modes of the $SU_L(2)$ gauge group on the off-shell $W^-W^+\gamma$ and $W^-W^+Z$ vertices is calculated in the context of a pure Yang-Mills theory in five dimensions and its phenomenological implications discussed. The use of a gauge-fixing procedure for the excited KK modes that is covariant under the standard gauge transformations of the $SU_L(2)$ group is stressed. A gauge-fixing term and the Faddeev-Popov ghost sector for the KK gauge modes that are separately invariant under the standard gauge transformations of $SU_L(2)$ are presented. It is shown that the one-loop contributions of the KK modes to the off shell $W^-W^+\gamma$ and $W^-W^+Z$ vertices are free of ultraviolet divergences and well-behaved at high energies. It is found that for a size of the fifth dimension of $R^{-1}\sim 1\, \mathrm{TeV}$, the one-loop contribution of the KK modes to these vertices is about one order of magnitude lower than the corresponding standard model radiative correction. This contribution is similar to the one estimated for new gauge bosons contributions in other contexts. Tree--level effects on these vertices induced by operators of higher canonical dimension are also investigated. It is found that these effects are lower than those generated at the one-loop order by the KK gauge modes.
\end{abstract}

\maketitle

\section{Introduction}
\label{Int}In the last decade, there has been considerable interest in studying the phenomenological implications of extra dimensions on low-energy observables, mainly since the pioneering works by Antoniadis, Arkani-Hamed, Dimopoulos and Dvali~\cite{antoniadis,AHDD,AADD}, where large extra dimensions were considered. In most scenarios, our observed 3-dimensional space is a 3-brane that is embedded in a higher $D$-dimensional space-time, which is known as the bulk. If the additional dimensions are small enough, the Standard Model (SM) gauge and matter fields are phenomenologically allowed to propagate in the bulk; otherwise they are stuck to the 3-brane. Of course, if there are extra dimensions, they must be smaller than the smallest scale which has been currently explored by experiments. So, the extra dimensions are assumed to be suitably compactified on some manifold of sufficiently small size. As a result of the compactification, the fields propagating in the bulk expand into series of states known as Kaluza-Klein (KK) towers, with the individual KK excitations being labeled by mode numbers. The collider signature for the existence of additional dimensions is the observation of a KK tower of states. While most of the studies have been restricted to tree--level processes, the quantum loop effects of the theory have received much less attention, as only some one-loop processes, as electromagnetic dipoles~\cite{ED}, the $b\to s\gamma$~\cite{BSG}, $Z\to \bar{b}b$~\cite{Papa,ZBB}, $B_{s,d}\to \gamma \gamma$~\cite{BsdGG} and $B_d\to l^{+}l^{-}$~\cite{Bdll} decays, including the contributions of virtual KK gravitons to the oblique parameters~\cite{TH}, and $B^{0}-\bar{B}^{0}$mixing~\cite{B0B0,ZBB}  have been considered. It is clear that any program that contemplates the calculation of radiative corrections of excited KK gauge modes on SM low-energy observables requires the introduction of a consistent quantization scheme for the four dimensional KK theory. In a recent publication by some of us, a consistent quantization scheme for the excited KK gauge modes of a pure $SU(N)$ theory in five dimensions, with the fifth dimension compactificated on the orbifold $S^1/Z_2$, was presented~\cite{NT}. As it was stressed in that work, to quantize the gauge KK modes it is necessary to identify the gauge transformations to which is subject the four dimensional theory, since the gauge parameters $\alpha^a(x,y)$ (with $y$ the fifth dimension) propagate in the bulk and thus their corresponding excited KK modes determine complicated nonstandard gauge transformations of the gauge KK modes. The precise identification of these new type of gauge transformations is crucial to quantize the theory on the basis of the Becchi--Rouet--Stora--Tyutin (BRST) symmetry~\cite{BRST}. In fact, the modern approach to the quantization of gauge systems based in the BRST symmetry requires to incorporate in the theory the gauge parameters as true degrees of freedom. As it was shown in Ref.~\cite{NT}, the four dimensional theory satisfies simultaneously the \textit{standard gauge transformations} (SGT) and one additional type of complicated gauge transformations, which we called \textit{nonstandard gauge transformations} (NSGT). As it is widely explained in this reference, the Yang-Mills fields in five-dimensions ${\cal A}^a_M(x,y)$ satisfy the SGT with parameters $\alpha^a(x,y)$. When the extra dimension is appropriately compactified and integrated, the Fourier series for ${\cal A}^a_\mu(x,y)$, ${\cal A}^a_5(x,y)$, and $\alpha^a(x,y)$ lead to infinite towers of KK modes. The zero modes of the fields ${\cal A}^a_\mu(x,y)$, denoted by $A^{(0)a}_\mu(x)$, are the gauge fields associated with the $SU_4(N)$ gauge group, whereas the zero modes of the $\alpha^a(x,y)$ parameters, $\alpha^{(0)a}$, determine the SGT of $SU_4(N)$. On the other hand, the variations defined by these zero-mode gauge parameters for the excited KK modes associated with the ${\cal A}^a_\mu(x,y)$ fields, $A^{(n)a}_\mu(x)$ ($n=1,2,\cdots $), as well as the ones arising from ${\cal A}^a_5(x,y)$, $A^{(n)a}_5(x)$, are transformations in the adjoint representation of $SU_4(N)$. It results that the excited KK modes of the gauge parameters, $\alpha^{(n)a}(x)$, determine the gauge transformations of the $A^{(n)a}_\mu(x)$ and $A^{(n)a}_5(x)$ KK modes. In contrast with the SGT, this new type of gauge invariance, cannot be easily identified. However, as it was emphasized in Ref.~\cite{NT}, the precise identification of these gauge transformations, as well as the covariant objects needed to construct invariants, is a first indispensable step in order to quantize the theory. The identification of this new type of gauge transformations, the construction of a classical action being invariant under both class of gauge transformations and its quantization on the basis of the BRST symmetry~\cite{BRST,PR}, as well as the introduction of a novel gauge-fixing procedure for the excited KK modes that is covariant under the SGT of $SU_4(N)$, are the main results of this reference.

There are phenomenological and theoretical motivations to quantize a gauge KK theory. If KK modes cannot be produced directly in the Large Hadron Collider (LHC) collider, it would be possible to detect their virtual effects through precision measurements as those planed to be realized in the International Linear Collider (ILC)~\cite{LHC-ILC}. Electroweak precision observables can play a role in various models. In many physics scenarios they can provide information about new physics scales that are too heavy to be detected directly. Due to this, it is crucial to count on a consistent quantum theory of the KK excitations that allows us to make predictions at the one-loop or higher orders. In particular, it is important to calculate the one-loop effects of these new particles on SM observables that eventually could be sensitive to new physics effects. Those processes firstly generated at the one-loop level within the SM are the best candidates, but the one-loop effects of KK modes on the trilinear $WW\gamma$ and $WWZ$ vertices will be of experimental interest in the ILC. On the theoretical side, it is interesting to investigate the behavior of the theory at the one-loop level. For instance, it is very important to study the UV structure of light Green functions, \textit{i.e.} Green functions consisting of zero modes only, due to one-loop contributions of excited modes. In fact, this is an important objective of this work.

The main goal of this work is to use the quantization scheme of reference~\cite{NT} within the context of the electroweak theory. In particular, we are interested in calculating the one-loop effects of the excited KK modes associated with the $SU_L(2)$ gauge group to the $W^-W^+W^3$ vertex, with $W^3$ an off shell $\gamma$ or $Z$ gauge boson. Apart from its phenomenological importance in the context of the ILC, this calculation will allow us to illustrate our quantization scheme for KK gauge modes~\cite{NT}, as we will show that the one-loop amplitude for this vertex, with $W^3$ off--shell, has an UV structure identical to the one generated by the radiative correction in the context of the SM. As it will be clarified below, this does not mean that contributions from two--loop or higher orders are well behaved at the UV domain. We will consider the case where the SM matter fields are rigidly fixed to the brane and do not feel the effects of the additional dimensions, which will be assumed flat. The simplest model of this class corresponds to gauge fields propagating in only one extra dimension. The gauge boson KK excitation masses are given by $m^2_n=(n/R)^2+m^2_0$, where $n$ labels the KK level, $R^{-1}$ is the compactification scale, and $m_0$ is the zero-mode mass, which is obtained via spontaneous symmetry breaking for the cases of $W$ and $Z$ and vanishes for $\gamma$. It is important to notice that the KK excitations of all the gauge states are highly degenerate, with a splitting too small to be observed at the LHC collider. Such a splitting will be still less important in radiative corrections, so it is feasible to assume that the KK modes are degenerate, with masses given by $n/R$.

Gauge models in more than four dimensions are nonrenormalizable\footnote{Aspects of nonperturbative renormalizability of gauge theories in more than four dimensions are analyzed in~\cite{NPR}.}, so they must be recognized as effective theories that become embedded in some other consistent UV completion, such as string theories. The nonrenormalizable nature of higher dimensional theories arises from the fact that they have dimensionfull constant couplings. So, the effective theory must be cut off at some scale $M_s$, above which the fundamental theory enters. To be specific, we will center our discussion by considering only one extra dimension. Although at the level of the four dimensional theory the coupling constants are dimensionless, the nonrenormalizable character manifest itself through the infinite multiplicity of the KK modes. Then, besides the Fermi scale $v\approx 246 \ \mathrm{GeV}$, the four dimensional effective theory have two additional scales, namely the compactification scale $R^{-1}$ and the cutoff $M_s$. The cutoff sensitivity of light Green's functions at the one-loop level depends on some aspects intimately connected with the compactified dimension. To see this, it should be noticed that any infinite dimension has associated, in the Fourier space, a continuous momentum, but a discrete momentum $k_5=n/R$ corresponds to the compactified coordinate. This means that a one-loop amplitude is determined by the usual continuous sum together with additional discrete sums: $\int d^4k \sum^{\infty}_{m,n,\ldots }$. At the one-loop level, only one discrete sum exists if the discrete momentum is conserved in each vertex. If this is the case, the KK parity, $(-1)^n$, is conserved, which means that no couplings involving only one single KK mode can arise. This in turns implies that no contributions to the electroweak observables can arise at the tree level~\cite{Apel}, although tree--level effects from operators of canonical dimension higher than $D(=5)$ can arise. So, at the one-loop level or higher orders two types of divergences can arise. Previous studies have shown~\cite{Papa,Apel} that in the so-called universal extra dimensional (UED) models, \textit{i.e.}, theories in which all the fields propagate in the extra dimension, the one-loop amplitudes for light Green's functions are insensitive to the cutoff $M_s$, as it remains only one discrete sum and it is convergent. In this paper, we will show that this is the case for the one-loop contributions of the gauge KK modes $W^{(m)}$ to the $WWW^3$ vertex. In this context of UED, when more than one extra dimension is introduced the one-loop effects of KK modes to light Green's functions cease to be independent on the cutoff~\cite{Apel}. On the other hand, in nonuniversal extra dimensional models, in which some fields are confined to the four dimensional brane, some new effects arise, as in this case the discrete momentum is not conserved in the brane but only in the bulk~\cite{Bdll}. In these models, vertices involving only one KK excited mode can exist and divergences can arise at the tree level, although the involved propagators are finite if only one extra dimension is considered~\cite{Bdll}. However, even in the case of nonuniversal exta dimensional models with only one extra dimension, one loop effects on light Green's functions are cutoff dependent~\cite{Bdll}. In this work, besides to show that the one-loop effects of the KK gauge modes on the $WWW^3$ vertex are insensitive to the cutoff, we will study the tree--level contribution to this vertex that arises from an operator of canonical dimension higher than $5$. It has been pointed out~\cite{Apel} that in UED models with only one extra dimension, the one-loop contribution on electroweak observables dominates~\cite{Apel}. We will see that this occurs in our case by showing that contributions induced by higher dimensional operators are lower than those induced by the KK modes at the one-loop level.

The rest of the paper has been organized as follows. Sec. \ref{Am} is devoted to calculate the one-loop contributions of the weak KK modes to the off shell $WWW^3$ vertex. The role of theoretical aspects as gauge-invariance and gauge-independence at the level of Green functions are discussed. The structure of the five dimensional $SU_5(2)$ theory and its compactification is presented. A gauge-fixing procedure for the excited KK modes that is covariant under the SGT of the $SU_L(2)$ group is introduced and the corresponding Faddeev-Popov ghost term derived. This section is concluded with the presentation of the form factors characterizing the one-loop amplitude for the $W^-W^+W^3$ vertex. In Sec. \ref{D}, we numerical results for the one-loop KK effects are discussed. Sec. \ref{hd} is devoted to discuss tree--level contributions to the $WWW^3$ vertex induced by operators of higher canonical dimension. The size of this contribution is compared with that generated at the one-loop by the KK excitations. Finally, in Sec. \ref{C} the conclusions are presented.

\section{One-loop effects of $SU_L(2)$ KK modes on the $WW\gamma$ and $WWZ$ couplings}
\label{Am}In this section, we apply the covariant gauge-fixing procedure derived in Ref.~\cite{NT} for investigating the one-loop impact of the gauge KK modes associated with the electroweak gauge bosons  to the off shell $WW\gamma$ and $WWZ$ vertices. As already commented in the introduction, there are two important motivations for studying these vertices in the context of KK theories. In first place is their intrinsic phenomenological interest in the context of precision measurements in future colliders. On the theoretical side, the study of radiative corrections to these vertices allow us to illustrate how our quantization scheme~\cite{NT} can be used in practical loop calculations. In particular, these calculations allow us to exemplify one of the main implications of this quantization scheme, \textit{i.e.} that it is possible to predict, in a consist way, one-loop effects of gauge KK modes on light Green functions.

\subsection{The off shell $W^-W^+W^3$ vertex: gauge invariance and gauge independence}

The radiative corrections to the  $WWW^3$ vertex ($W^3=\gamma\, , Z$) have been the subject of considerable interest in the literature. Apart from its sensitivity to new physics effects, this vertex has theoretical interest of its own as it may serve as a probe of the gauge sector of the SM. In this context, the one-loop contributions to the on-shell  $WW\gamma$ vertex, which defines the static electromagnetic properties of the W boson, have been calculated in the SM~\cite{WWgSM} and most of its extensions such as two-Higgs doublet models~\cite{WWgTHDM}, supersymmetric models~\cite{WWgSUSY}, composite particles~\cite{WWgComposite}, models with an extra $Z'$ boson~\cite{WWgZ'}, left-right symmetric models~\cite{WWgLR}, $331$ models~\cite{WWg331}, the littlest Higgs model\cite{LHM}, and in a model independent way using an effective Lagrangian approach~\cite{ELA,WWgEL}. Although the study of the one-loop on-shell $WW\gamma$ vertex is important to quantify the impact of new physics effects, they are the $WW\gamma$ and $WWZ$ vertices, with $\gamma$ and $Z$ off-shell, the ones which are of interest for investigating the one-loop effects of new particles and thus to have indirect evidence of their presence through precision measurements at high energy collisions, as those planed for the $e^+e^-$ International Linear Collider~\cite{ILC}. In the context of a CP conserving theory, the one-loop radiative corrections to the $WWW^3$ vertex can be parametrized by two form factors, $\Delta \kappa_{W^3}$ and $\lambda_{W^3}$, which for the especial case of $W^3=\gamma$ on shell define the CP-even electromagnetic properties of the $W$ gauge boson, namely, the magnetic dipole moment and the electric quadrupole moment~\cite{WWgSM,Hagiwara}.

 The calculation of radiative corrections to the $WWW^3$ vertex arising from new gauge bosons, as the KK modes, must be treated with some care when at least one of the three particles of this coupling is off-shell, as, under such circumstances, this calculation is sensitive to the gauge-fixing procedure used to define the propagators of the gauge fields circulating in the loop. This contrasts with the on-shell $WW\gamma$ vertex, which corresponds to a $S$-matrix element and is, therefore, a gauge-independent quantity. It is well-known that the conventional gauge-fixing procedures~\cite{CGFP} give rise to ill-behaved off-shell Green functions that may display inadequate properties such as a nontrivial dependence on the gauge-fixing parameter, an increase larger than the one observed in physical amplitudes at high energies, and the appearance of unphysical thresholds. It would be interesting if one was able to study the sensitivity to radiative corrections of the $WWW^3$ coupling without invoking some particular $S$-matrix element. Behind this are the concepts of gauge invariance and gauge independence, which are essential ingredients of the gauge systems. Although the former play a fundamental role to define the classical action, it does not survive to quantization, as one must invariably invoke an appropriate gauge-fixing procedure to define the quantum action. At the quantum level, the theory is governed by the BRST symmetry~\cite{BRST,PR}, which generates Green functions satisfying the Slavnov-Taylor identities instead of the simpler ones that would exist if the quantum action was gauge-invariant. Fortunately, there are other gauge-fixing procedures that allow us to construct well-behaved off shell Green functions. One of these nonconventional gauge-fixing procedures is the background field method (BFM)~\cite{BFM}, which allows one to construct a quantum action  satisfying a sort of gauge invariance. The method consists in unfolding the gauge fields, $A^a_\mu$, into a quantum, $Q^a_\mu$, and a classical, $\hat{A}^a_\mu$, parts: $A^a_\mu \rightarrow \hat{A}^a_\mu+Q^a_\mu$. While the effective quantum action is defined through the path integral on the $Q^a_\mu$ fields, the classical fields $\hat{A}^a_\mu$ play the role of sources with respect to which the vertex functions are derived. Due to this, it is only necessary to introduce a gauge-fixing procedure for the quantum fields $Q^a_\mu$ and thus the resultant quantum theory is invariant under gauge transformations of the background fields $\hat{A}^a_\mu$. The Green functions derived in this context satisfy simple (QED-like) Ward identities, which are well-behaved because they contain less unphysical information in comparison with those that arise from the conventional quantization methods. However, it is worth stressing that they are still dependent on the gauge parameter $\xi_Q$ that characterizes the gauge-fixing scheme used for the quantum fields, and so there is no gauge independence. Although gauge dependent, one expects that these Green functions provide us information quite near to the physical reality. The BFM has proved to be useful in many applications~\cite{ABFM,C1L}, simplifying both technically and conceptually the calculation of radiative corrections. At the present, there is still no known mechanism that allows one to construct a quantum action from which can be derived both gauge-invariant and gauge-independent Green functions, although there is already a diagrammatic method meant for this purpose, the so-called pinch technique (PT)~\cite{PT,PRPT,PTSSB}. This method consists in constructing well-behaved Green functions of a given number of points by combining some individual contributions from Green functions of equal and higher number of points, whose Feynman rules are derived from a conventional effective action or even from a nonconventional scheme as the BFM~\cite{BFMPT}. The PT has been used successfully in pure Yang-Mills theories~\cite{PT,BFMPT,PTYM,PTSSB} as well as in theories with spontaneous symmetry breaking (SSB)~\cite{PTSSB}. An important application has been the study of self-energies~\cite{PTSE} and trilinear vertices~\cite{PTTV} involving the electroweak gauge bosons. In particular, a complete calculation of the one-loop  $WWW^3$ vertex in the context of the electroweak theory showed that a simple Ward identity is satisfied by this vertex and the $W$ self-energy~\cite{PTWWV1,PTWWV2}. The gauge independence of the $WW\gamma$ vertex for off-shell photon and on-shell $W$ bosons is discussed in Ref.~\cite{PTWWV1}. There is a remarkable connection between the PT and the BFM which consists in the fact that the Green functions calculated via the BFM Feynman rules coincide with those obtained through the PT for the specific value $\xi_Q=1$. This interesting property was first established at the one-loop level~\cite{C1L}, next confirmed at the two-loop level~\cite{C2L}, and more recently at any order of perturbation theory~\cite{CANYL}. The reason for such a link remains a puzzle, though it is worth noting that the Feynman-'t Hooft gauge yields no unphysical thresholds.

Although in conventional quantization schemes the quantum action of the theory is not gauge-invariant, it is still possible to introduce gauge invariance with respect to a subgroup of such a theory. This scheme is particularly useful to assess the virtual effects of heavy gauge bosons lying beyond the Fermi scale on the SM Green functions in a $SU_L(2)\times U_Y(1)$-covariant manner, in which case it is only necessary to introduce a quantization scheme for the heavy fields since the SM fields would only appear as external legs. A scheme of this class was proposed by some of us some few years ago~\cite{CG331} to investigating the loop effects of new heavy gauge bosons predicted by the so-called $331$ models~\cite{PF} on the off-shell $WWW^3$ vertex. This model, which is based on the $SU_C(3)\times SU_L(3)\times U_X(1)$ gauge group, predicts the existence of five new heavy gauge bosons, two doubly charged, two simply charged, and a neutral $Z'$~\cite{Z331}. These particles acquire their masses in the first stage of SSB, when the $SU_L(3)\times U_X(1)$ group is broken down into the usual electroweak group $SU_L(2)\times U_Y(1)$. At this scale, the charged gauge bosons arise in the fundamental representation of $SU_L(2)$, \textit{i.e.}, they appear as the components of the doublet  $Y^\dag_\mu=(Y^{--}_\mu,Y^-_\mu)$, which has hypercharge $Y=3$. In Ref.~\cite{CG331}, a gauge-fixing procedure for the $Y_\mu$ doublet of gauge bosons was introduced in a covariant way under the $SU_L(2)\times U_Y(1)$ group. Following the procedure used by Fujikawa, in the context of the SM~\cite{FNLG}, to define the $W$ propagator in a covariant way under the electromagnetic $U_e(1)$ group, nonlinear gauge-fixing functions transforming in the same way that $Y_\mu$ does under the electroweak group were introduced~\cite{CG331}. Well-behaved amplitudes for the $WW\gamma $ and $WWZ$ vertices, with $\gamma$ and $Z$ off-shell, were derived in a simple way as a consequence of the high symmetry of the gauge-fixing procedure. This gauge-fixing procedure is quite similar to the one presented in Ref.~\cite{NT} for the KK modes, which is covariant under the SGT of $SU_4(N)$. In the present work, we will use this gauge-fixing procedure for the gauge KK modes  $W^{(n)\pm}$ and $W^{(n)3}$ associated with the $SU_L(2)$ group. Since we are only interested in the one-loop effects of the KK modes on the light Green function $W^-W^+W^3$, we do not need to fix the gauge for the zero KK modes $W^{(0)\pm}\equiv W^\pm$ and $W^{(0)3}\equiv W^3$. This means that the gauge-fixed quantum action is invariant under the SGT of the $SU_L(2)$ group, which in turn implies that the $W^-W^+W^3$ coupling will satisfy simple Ward identities. Although, as already commented, the one-loop amplitudes are still gauge-dependent, they are well-behaved if calculated in the Feynman-'t Hooft gauge.

\subsection{The $SU_L(2)$ Yang-Mills sector in five dimensions}
The structure of the four dimensional effective Lagrangian for the $SU_L(2)$ KK theory is identical to the one given in Ref.~\cite{NT} for the compactified theory based in the $SU_4(N)$ group. Due to this, we will present only the essential steps of its derivation. The five dimensional theory is characterized by the following gauge invariant action
\begin{equation}
S=\int d^4x\int dy\left(-\frac{1}{4}{\cal W}^a_{MN}(x,y){\cal W}^{MN}_a(x,y) \right)\, ,
\end{equation}
where the five dimensional strength tensor is given by
\begin{equation}
{\cal W}^a_{MN}=\partial_M {\cal W}^a_N-\partial_N {\cal W}^a_M+g_5 \epsilon^{abc}{\cal W}^b_M{\cal W}^c_N\,.
\end{equation}
In the following, we will denote by $x$ the usual four coordinates and by $y$ the fifth dimension. We will employ a flat metric with signature $diag(1,-1,-1,-1,-1)$. In addition, $M=0,1,2,3,5$ and $a=1,2,3$ stand for Lorentz and gauge indices, respectively. The following periodicity and parity properties of the gauge fields and gauge parameters are assumed:
\begin{eqnarray}
{\cal W}^a_{MN}(x,y+2\pi R)&=&{\cal W}^a_{MN}(x,y)\, ,\\
{\cal W}^a_{M}(x,y+2\pi R)&=&{\cal W}^a_{M}(x,y)\, ,\\
\alpha^a(x,y+2\pi R)&=&\alpha^a(x,y)\, ,
\end{eqnarray}
\begin{eqnarray}
{\cal W}^a_{\mu \nu}(x,-y)&=&{\cal W}^a_{\mu \nu}(x,y)\, ,\\
{\cal W}^a_{\mu 5}(x,-y)&=&-{\cal W}^a_{\mu 5}(x,y)\, \\
{\cal W}^a_{\mu}(x,-y)&=&{\cal W}^a_{\mu}(x,y)\, ,\\
{\cal W}^a_5(x,-y)&=&-{\cal W}^a_5(x,y)\, \\
\alpha^a(x,-y)&=&\alpha^a(x,y)\, ,
\end{eqnarray}
where the Greek indices run, as usual, from $0$ to $3$. This allows us to express the strength tensor, the gauge fields, and the gauge parameters as Fourier series:
\begin{eqnarray}
{\cal W}^a_{\mu \nu}(x,y)&=&\frac{1}{\sqrt{2\pi R}}{\cal W}^{(0)a}_{\mu \nu}(x)+\sum^\infty_{m=1}\frac{1}{\sqrt{\pi R}}{\cal W}^{(m)a}_{\mu \nu}(x)\cos\left(\frac{my}{R}\right)\, ,\\
{\cal W}^a_{\mu 5}(x,y)&=&\sum^\infty_{m=1}\frac{1}{\sqrt{\pi R}}{\cal W}^{(m)a}_{\mu 5}(x)\sin\left(\frac{my}{R}\right)\, ,
\end{eqnarray}
\begin{eqnarray}
{\cal W}^a_{\mu }(x,y)&=&\frac{1}{\sqrt{2\pi R}}W^{(0)a}_{\mu}(x)+\sum^\infty_{m=1}\frac{1}{\sqrt{\pi R}}W^{(m)a}_{\mu }(x)\cos\left(\frac{my}{R}\right)\, ,\\
{\cal W}^a_{5}(x,y)&=&\sum^\infty_{m=1}\frac{1}{\sqrt{\pi R}}W^{(m)a}_{5}(x)\sin\left(\frac{my}{R}\right)\, ,
\end{eqnarray}
\begin{equation}
\alpha^a(x,y)=\frac{1}{\sqrt{2\pi R}}\alpha^{(0)a}(x)+\sum^\infty_{m=1}\frac{1}{\sqrt{\pi R}}\alpha^{(m)a}(x)\cos\left(\frac{my}{R}\right)\, .
\end{equation}
To preserve gauge invariance, the fifth dimension must be integrated in the action $S$ by considering only covariant objects~\cite{NT}, \textit{i.e.} by expanding in Fourier series the strength tensors instead of the gauge fields,
\begin{eqnarray}
{\cal L}^{SU_L(2)}_{4{\rm YM}}&=&-\frac{1}{4}\int^{2\pi R}_0 \left[{\cal W}^a_{\mu \nu}(x,y){\cal W}^{\mu \nu}_a(x,y)+{\cal W}^a_{\mu 5}(x,y){\cal W}^{\mu 5}_a(x,y)\right] \nonumber \\
&=&-\frac{1}{4}\left({\cal W}^{(0)a}_{\mu \nu}{\cal W}^{(0)a \mu \nu}+{\cal W}^{(m)a}_{\mu \nu}{\cal W}^{(m)a \mu \nu}\right)+\frac{1}{2}{\cal W}^{(m)a}_{\mu 5}{\cal W}^{(m)a\mu}\hspace{0.01cm}_5\, ,
\label{eq:4DeffL}
\end{eqnarray}
where sums over all type of repeated indices, including the Fourier ones, are assumed. This convention will be maintained through the paper. In the above expressions,
\begin{equation}
{\cal W}^{(0)a}_{\mu \nu}=W^{(0)a}_{\mu \nu}+g\epsilon^{abc}W^{(m)b}_\mu W^{(m)c}_\nu\, ,
\end{equation}
\begin{equation}
{\cal W}^{(m)a}_{\mu \nu}={\cal D}^{(0)ab}_\mu W^{(m)b}_\nu-{\cal D}^{(0)ab}_\nu W^{(m)b}_\mu+ g\epsilon^{abc}\Delta^{mrn}W^{(r)b}_\mu W^{(n)c}_\nu \, ,
\end{equation}
\begin{equation}
{\cal W}^{(m)a}_{\mu 5}={\cal D}^{(0)ab}_\mu W^{(m)b}_5+\frac{m}{R}W^{(m)a}_\mu +g\epsilon^{abc}\Delta'^{mrn}W^{(r)b}_\mu W^{(n)c}_5\, ,
\end{equation}
where
\begin{equation}
W^{(0)a}_{\mu \nu}=\partial_\mu W^{(0)a}_\nu-\partial_\nu W^{(0)a}_\mu +g\epsilon^{abc}W^{(0)b}_\mu W^{(0)c}_\nu \, ,
\end{equation}
is the standard strength tensor of $SU_L(2)$ and ${\cal D}^{(0)ab}_\mu=\delta^{ab}\partial_\mu-g\epsilon^{abc}W^{(0)c}_\mu$ is the covariant derivative in the adjoint representation of this group. In addition,
\begin{eqnarray}
\Delta^{kmn}&=&\frac{1}{\sqrt{2}}\left(\delta_{k,m+n} +\delta_{m,k+n}+\delta_{n,k+m}\right)\, , \\
\Delta'^{kmn}&=&\frac{1}{\sqrt{2}}\left(\delta_{k,m+n} +\delta_{m,k+n}-\delta_{n,k+m}\right)\, .
\end{eqnarray}
The above four dimensional Lagrangian, equation \ref{eq:4DeffL}, is invariant under the SGT of $SU_L(2)$~\cite{NT}, in which $W^{(0)a}_\mu$ transforms as a gauge field, whereas the excited modes $W^{(m)a}_\mu$ and $W^{(m)a}_5$ transform as matter fields in the adjoint representation of this group:
\begin{eqnarray}
\delta W^{(0)a}_\mu&=&{\cal D}^{(0)ab}_\mu \alpha^{(0)b}\, ,\\
\delta W^{(m)a}_\mu&=&g\epsilon^{abc}W^{(m)b}_\mu \alpha^{(0)c} \, , \\
\delta W^{(m)a}_5&=&g\epsilon^{abc}W^{(m)b}_5\alpha^{(0)c}\, .
\end{eqnarray}
This symmetry is manifest in the ${\cal L}^{SU_L(2)}_{4 {\rm YM}}$ Lagrangian. Although less evident, this Lagrangian is also invariant under the following NSGT~\cite{NT},
\begin{eqnarray}
\delta W^{(0)a}_\mu&=&g\epsilon^{abc}W^{(m)b}\alpha^{(m)c}\, ,\\
\delta W^{(m)a}_\mu&=&{\cal D}^{(mn)ab}_\mu \alpha^{(n)b}\, ,\\
\delta W^{(m)a}_5&=&D^{(mn)ab}_5\alpha^{(n)b}\, ,
\end{eqnarray}
where
\begin{equation}
{\cal D}^{(mn)ab}_\mu=\delta^{mn}{\cal D}^{(0)ab}_\mu-g\epsilon^{abc}\Delta^{mrn}W^{(r)c}_\mu \, ,
\end{equation}
\begin{equation}
D^{(mn)ab}_5=-\delta^{mn}\delta^{ab}\frac{m}{R}-g\epsilon^{abc}\Delta'^{mrn}W^{(r)c}_5\, .
\end{equation}
The strength tensors ${\cal W}^{(0)a}_{\mu \nu}$, ${\cal W}^{(m)a}_{\mu \nu}$, and ${\cal W}^{(m)a}_{\mu 5}$ transform in a well-defined way under both the SGT and the NSGT~\cite{NT} and thus ${\cal L}^{SU_L(2)}_{4 {\rm YM}}$ is gauge invariant.

\subsection{The gauge-fixing procedure}

As already commented, it is very important from the phenomenological point of view to investigate the one-loop impact of the excited KK modes on light Green functions because their sensitivity to these virtual effects of new physics could in principle be confronted with precision measurements that will be realized at future linear colliders. To calculate such virtual effects, a gauge-fixing procedure that allows us to define the propagators of the excited KK modes must be implemented. In Ref.~\cite{NT}, a gauge-fixing procedure for the excited KK modes based on the BRST symmetry, and which is covariant under the SGT of $SU_4(N)$, was introduced and the corresponding ghost sector derived. In this work, we only present the main ingredients of such a quantization scheme. The degeneration of the theory due to the presence of the NSGT is removed via the following gauge-fixing functions
\begin{equation}
f^{(m)a}={\cal D}^{(0)ab}_\mu W^{(m)b \mu}-\xi \frac{m}{R}W^{(m)a}_5\, .
\end{equation}
Notice that these functions transform in the adjoint representation of $SU_L(2)$, and therefore they lead to a quantized theory that preserves gauge invariance with respect to the SGT of $SU_L(2)$.
It is worth emphasizing that our gauge-fixing approach permits one to fix the gauge for the zero modes and for the excited ones, independently of each other. Indeed, the fixation for the zero modes can be performed as in the standard four dimensional Yang-Mills theory.
Since we are only interested in quantifying the one-loop effects of the excited KK modes on light Green functions, we do not need to introduce a gauge-fixing procedure for the zero modes $W^{(0)a}_\mu$ and thus, at this stage, the SGT are preserved at the quantum level. The corresponding gauge-fixing Lagrangian is given by
\begin{equation}
{\cal L}_{\rm GF}=-\frac{1}{2\xi}\left({\cal D}^{(0)ab}_\mu W^{(m)b\mu}\right)\left({\cal D}^{(0)ac}_\nu W^{(m)c\nu} \right)+m_m W^{(m)a}_5\left({\cal D}^{(0)ab}_\mu W^{(m)b\mu}\right)-\frac{1}{2}\xi m^2_m W^{(m)a}_5W^{(m)a}_5\, ,
\end{equation}
where $m_m=(m/R)$ is the mass of the excited KK mode $W^{(m)}_\mu$ and $\sqrt{\xi}m_m$ the mass of the associated pseudo Goldstone boson $W^{(m)a}_5$. This gauge-fixing procedure allows us to cancel the bilinear and trilinear nonphysical couplings $W^{(m)a}_\mu W^{(m)b}_5$ and $W^{(0)a}_\mu W^{(m)b}_\nu W^{(m)c}_5$ that are present in the  ${\cal L}^{SU_L(2)}_{4 {\rm YM}}$ Lagrangian through a total derivative:
\begin{eqnarray}
{\cal L}_{\rm GF}+\frac{1}{2}{\cal W}^{(m)a}_{\mu 5}{\cal W}^{(m)a\mu}\hspace{0.01cm}_5&=&m_m\left[ W^{(m)a}_5\left({\cal D}^{(0)ab}_\mu W^{(m)b\mu}\right)+W^{(m)a\mu}\left( {\cal D}^{(0)ab}_\mu W^{(m)b}_5 \right)\right]+\cdots \nonumber \\
&=&m_m\partial_\mu \left(W^{(m)a}_5 W^{(m)a\mu} \right)+\cdots
\end{eqnarray}

On the other hand, the ghost sector induced by this gauge-fixing procedure can be written as the sum of two terms~\cite{NT},
\begin{equation}
{\cal L}_{\rm FP}={\cal L}^1_{\rm FP}+{\cal L}^2_{\rm FP}\, ,
\end{equation}
where
\begin{equation}
{\cal L}^1_{\rm FP}=\bar{C}^{(m)c}\left({\cal D}^{(0)ac}_\mu {\cal D}^{(mn)ab\mu}+\xi m_m D^{(mn)cb}_5 \right)C^{(n)b}-\xi g\epsilon^{abc}\Delta^{mrn}f^{(m)a}\bar{C}^{(r)b}C^{(n)c} \, .
\end{equation}
The term ${\cal L}^2_{\rm FP}$ is not relevant for our purposes, as it involves only quartic interactions among ghost and antighost fields~\cite{NT} and we do not present it here. It is important to notice that the Lagrangian ${\cal L}^1_{\rm FP}$ is invariant under the SGT of $SU_L(2)$.

It is worth presenting the pieces of the four dimensional theory that can contribute, at the one-loop order, to light Green functions:
\begin{equation}
-\frac{1}{4}{\cal W}^{(0)a}_{\mu \nu}{\cal W}^{(0)a\mu \nu}\rightarrow -\frac{1}{2}g \epsilon^{abc}W^{(0)a}_{\mu \nu}W^{(n)b\mu}W^{(n)c\nu} \, ,
\end{equation}
\begin{equation}
-\frac{1}{4}{\cal W}^{(n)a}_{\mu \nu}{\cal W}^{(n)a\mu \nu} \rightarrow \frac{1}{4}\left( {\cal D}^{(0)ab}_\mu W^{(n)b}_\nu-{\cal D}^{(0)ab}_\nu W^{(n)b}_\mu\right)\left( {\cal D}^{(0)ac\mu}W^{(n)c\nu}-{\cal D}^{(0)ac\nu}W^{(n)c\mu}\right)\, ,
\end{equation}
\begin{equation}
\frac{1}{2}{\cal W}^{(n)a}_{\mu 5}{\cal W}^{(n)a\mu}_5\rightarrow \frac{1}{2}\left({\cal D}^{(0)ab}_\mu W^{(n)b}_5 \right)\left({\cal D}^{(0)ac\mu}W^{(n)c}_5 \right)+\frac{1}{2}m^2_nW^{(n)a}_\mu W^{(n)a\mu}\, ,
\end{equation}
\begin{equation}
{\cal L}_{\rm GF}\rightarrow -\frac{1}{2\xi}\left({\cal D}^{(0)ab}_\mu W^{(n)b\mu} \right)\left({\cal D}^{(0)ac}_\nu W^{(n)c\nu} \right)-\frac{1}{2}\xi m^2_n W^{(n)a}_5W^{(n)a}_5\, ,
\end{equation}
\begin{equation}
{\cal L}_{\rm FP}^1\rightarrow \bar{C}^{(n)b}{\cal D}^{(0)ab}_\mu {\cal D}^{(0)ac\mu}C^{(n)c}-\xi m^2_n \bar{C}^{(n)a}C^{(n)a}\, .
\end{equation}

\subsection{Feynman rules}
We now turn to list the vertices needed to calculate the contribution of the excited KK modes to the off-shell $W^-W^+W^3$ vertex. To do this, we introduce the physical basis:
\begin{equation}
W^{\pm}_\mu=\frac{1}{\sqrt{2}}\left(W^1_\mu \mp iW^2_\mu \right)\, ,
\end{equation}
\begin{equation}
W^{(n)\pm}_\mu=\frac{1}{\sqrt{2}}\left(W^{(n)1}_\mu \mp iW^{(n)2}_\mu \right)\, .
\end{equation}
There are trilinear and quartic vertices that can contribute at the one--loop level to the $W^-W^+W^3$ coupling. The trilinear gauge couplings are given by
\begin{eqnarray}
{\cal L}_{W^3W^{(n)-}W^{(n)+}}&=&-ig\Big[\left(W^{(n)+}_{\mu \nu}W^{(n)-\nu}-W^{(n)-}_{\mu \nu}W^{(n)+\nu} \right)W^{3\mu}+W^3_{\mu \nu}W^{(n)-\mu}W^{(n)+\nu} \nonumber \\
&& -\frac{1}{\xi}W^3_\mu\left(W^{(n)+\mu}\partial_\nu W^{(n)-\nu}-W^{(n)-\mu}\partial_\nu W^{(n)+\nu} \right)\Big]\, ,
\end{eqnarray}
\begin{eqnarray}
{\cal L}_{W^+W^{(n)-}W^{(n)3}}&=&-ig\Big[\left(W^{(n)-}_{\mu \nu}W^{(n)3\nu}-W^{(n)3}_{\mu \nu}W^{(n)-\nu} \right)W^{+\mu}+W^+_{\mu \nu}W^{(n)3\mu}W^{(n)-\nu}\nonumber \\
&&-\frac{1}{\xi}W^+_\mu\left(W^{(n)-\mu}\partial_\nu W^{(n)3\nu}-W^{(n)3\mu}\partial_\nu W^{(n)-\nu} \right)  \Big]\, ,
\end{eqnarray}

\begin{eqnarray}
{\cal L}_{W^-W^{(n)+}W^{(n)3}}&=&ig\Big[\left(W^{(n)+}_{\mu \nu}W^{(n)3\nu}-W^{(n)3}_{\mu \nu}W^{(n)+\nu} \right)W^{-\mu}+W^-_{\mu \nu}W^{(n)3\mu}W^{(n)+\nu}\nonumber \\
&&-\frac{1}{\xi}W^-_\mu\left(W^{(n)+\mu}\partial_\nu W^{(n)3\nu}-W^{(n)3\mu}\partial_\nu W^{(n)+\nu} \right)  \Big]\, ,
\end{eqnarray}
where $V_{\mu \nu}=\partial_\mu V_\nu-\partial_\nu V_\mu$, with $V$ standing for $W^\pm\, , W^{(n)\pm}\, , W^3\,,$ and $ W^{(n)3}$. On the other hand, the quartic vertices that can contribute to light Green functions, at the one-loop level, are:
\begin{equation}
{\cal L}_{W^3W^3W^{(n)-}W^{(n)+}}=-g^2\left[W^3_\mu W^{(n)+}_\nu\left(W^{3\mu}W^{(n)-\nu}-W^{3\nu}W^{(n)-\mu} \right)+\frac{1}{\xi}W^3_\mu W^3_\nu W^{(n)+\mu}W^{(n)-\nu} \right]\, ,
\end{equation}
\begin{eqnarray}
{\cal L}_{W^3W^+W^{(n)3}W^{(n)-}}&=&-g^2\Big[W^3_\mu W^{(n)-}_\nu\left(W^{(n)3\mu}W^{+\nu}-W^{(n)3\nu}W^{+\mu} \right)\nonumber \\
&&+W^{(n)3\mu}W^{(n)-\nu}\left(W^3_\mu W^+_\nu-W^3_\nu W^+_\mu \right)-\frac{1}{\xi}W^{(n)3\mu}W^3_\nu W^{(n)-\nu}W^+_\mu  \Big]\, ,
\end{eqnarray}
\begin{eqnarray}
{\cal L}_{W^3W^-W^{(n)3}W^{(n)+}}&=&-g^2\Big[W^3_\mu W^{(n)+}_\nu\left(W^{(n)3\mu}W^{-\nu}-W^{(n)3\nu}W^{-\mu} \right)\nonumber \\
&&+W^{(n)3\mu}W^{(n)+\nu}\left(W^3_\mu W^-_\nu-W^3_\nu W^-_\mu \right)-\frac{1}{\xi}W^{(n)3\mu}W^3_\nu W^{(n)+\nu}W^-_\mu  \Big]\, .
\end{eqnarray}
\begin{equation}
{\cal L}_{W^-W^+W^{(n)3}W^{(n)3}}=-g^2\Big[W^{(n)3}_\mu W^+_\nu\left(W^{(n)3\mu}W^{-\nu}-W^{(n)3\nu}W^{-\mu} \right)+\frac{1}{\xi}W^-_\mu W^+_\nu W^{(n)3\mu}W^{(n)3\nu} \Big]\, ,
\end{equation}
\begin{eqnarray}
{\cal L}_{W^-W^+W^{(n)-}W^{(n)+}}&=&g^2\Big[W^+_\mu W^{(n)-}_\nu\left(W^{-\nu}W^{(n)+\mu}-W^{-\mu}W^{(n)+\nu} \right)\nonumber \\
&&+W^{(n)-}_\mu W^{(n)+}_\nu\left(W^{-\mu}W^{+\nu}-W^{-\nu}W^{+\mu} \right) -\frac{1}{\xi}W^+_\mu W^-_\nu W^{(n)-\mu}W^{(n)+\nu}\Big]\, ,
\end{eqnarray}
\begin{equation}
{\cal L}_{W^+W^+W^{(n)-}W^{(n)-}}=\frac{g^2}{2}\Big[W^+_\mu W^{+\mu}W^{(n)-}_\nu W^{(n)-\nu}-\left(1-\frac{1}{\xi}\right)W^+_\mu W^+_\nu W^{(n)-\mu}W^{(n)-\nu} \Big]\, ,
\end{equation}
\begin{equation}
{\cal L}_{W^-W^-W^{(n)+}W^{(n)+}}=\frac{g^2}{2}\Big[W^-_\mu W^{-\mu}W^{(n)+}_\nu W^{(n)+\nu}-\left(1-\frac{1}{\xi}\right)W^-_\mu W^-_\nu W^{(n)+\mu}W^{(n)+\nu} \Big]\, ,
\end{equation}

The trilinear and quartic couplings among pseudo--Goldstone bosons and gauge fields that can contribute, at the one--loop level, to light Green functions are given by
\begin{equation}
{\cal L}_{W^3W^{(n)-}_5W^{(n)+}_5}=igW^3_\mu\left(W^{(n)-}_5\partial^\mu W^{(n)+}_5-W^{(n)+}_5\partial^\mu W^{(n)-}_5 \right)\, ,
\end{equation}
\begin{equation}
{\cal L}_{W^+W^{(n)-}_5W^{(n)3}_5}=igW^+_\mu\left(W^{(n)3}_5\partial^\mu W^{(n)-}_5-W^{(n)-}_5\partial^\mu W^{(n)3}_5 \right)\, ,
\end{equation}
\begin{equation}
{\cal L}_{W^-W^{(n)+}_5W^{(n)3}_5}=-igW^-_\mu\left(W^{(n)3}_5\partial^\mu W^{(n)+}_5-W^{(n)+}_5\partial^\mu W^{(n)3}_5 \right)\, ,
\end{equation}
\begin{eqnarray}
{\cal L}_{W^3W^3W^{(n)-}_5W^{(n)+}_5}&=&g^2W^3_\mu W^{3\mu} W^{(n)-}_5W^{(n)+}_5\, , \\ \nonumber \\
{\cal L}_{W^-W^+W^{(n)-}_5W^{(n)+}_5}&=&g^2W^-_\mu W^{+\mu} W^{(n)-}_5W^{(n)+}_5\, , \\ \nonumber \\
{\cal L}_{W^3W^{+}W^{(n)-}W^{(n)3}}&=&-g^2W^3_\mu W^{+\mu} W^{(n)-}_5W^{(n)3}_5\, , \\ \nonumber \\
{\cal L}_{W^3W^{-}W^{(n)+}W^{(n)3}}&=&-g^2W^3_\mu W^{-\mu} W^{(n)+}_5W^{(n)3}_5\, .
\end{eqnarray}

Finally, the part of the ghost sector that can contribute, at the one-loop level, to light Green functions can be written as follows:
\begin{eqnarray}
{\cal L}_{WC\bar{C}}&=&-ig\Big[W^3_\mu\left(C^{(n)+}\partial^\mu \bar{C}^{(n)-}-\partial^\mu C^{(n)+}\bar{C}^{(n)-} -C^{(n)-}\partial^\mu \bar{C}^{(n)+}+\partial^\mu C^{(n)-}\bar{C}^{(n)+}\right) \nonumber \\
&&-W^+_\mu\left(C^{(n)3}\partial^\mu \bar{C}^{(n)-}-\partial^\mu C^{(n)3}\bar{C}^{(n)-}+\partial^\mu C^{(n)-}\bar{C}^{(n)3}-C^{(n)-}\partial^\mu \bar{C}^{(n)3} \right)\nonumber \\
&&+W^-_\mu\left(C^{(n)3}\partial^\mu \bar{C}^{(n)+}-\partial^\mu C^{(n)3}\bar{C}^{(n)+}+\partial^\mu C^{(n)+}\bar{C}^{(n)3}-C^{(n)+}\partial^\mu \bar{C}^{(n)3} \right)\Big]\, ,
\end{eqnarray}
\begin{eqnarray}
{\cal L}_{WWC\bar{C}}&=&g^2\Big[W^-_\mu W^{+\mu}\left(C^{(n)+}\bar{C}^{(n)-}+C^{(n)-}\bar{C}^{(n)+}+2C^{(n)3}\bar{C}^{(n)3} \right)\nonumber \\
 &&-W^+_\mu W^{+\mu}C^{(n)-}\bar{C}^{(n)-}-W^-_\mu W^{-\mu}C^{(n)+}\bar{C}^{(n)+}-W^3_\mu W^{3\mu}C^{(n)3}\bar{C}^{(n)3}\nonumber \\
 &&-W^3_\mu\left( W^{+\mu}\left(C^{(n)-}\bar{C}^{(n)3}+C^{(n)3}\bar{C}^{(n)-} \right)+W^{-\mu}\left(C^{(n)+}\bar{C}^{(n)3}+C^{(n)3}\bar{C}^{(n)+} \right)\right)\Big]\, ,
\end{eqnarray}
where we have introduced the definitions
\begin{equation}
C^{(n)\pm}=\frac{1}{\sqrt{2}}\left(C^{(n)1}\mp iC^{(n)2} \right)\, ,
\end{equation}
\begin{equation}
\bar{C}^{(n)\pm}=\frac{1}{\sqrt{2}}\left(\bar{C}^{(n)1}\mp i\bar{C}^{(n)2} \right)\, .
\end{equation}

\begin{figure}[!ht]
\centering
\includegraphics[width=3.5in]{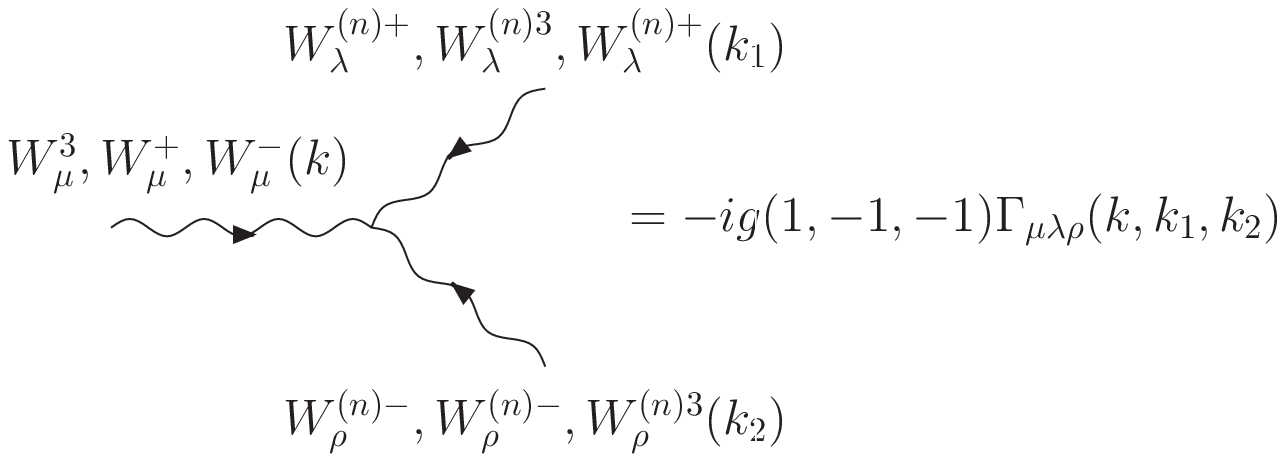} \ \ \ \
\includegraphics[width=2.5in]{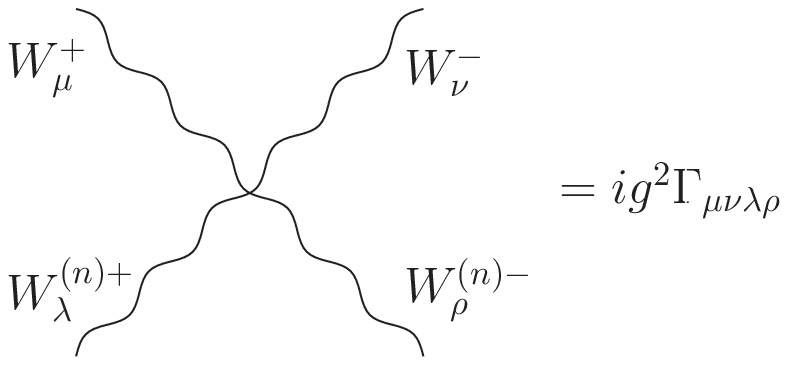}
\vskip 0.5cm
\includegraphics[width=2.5in]{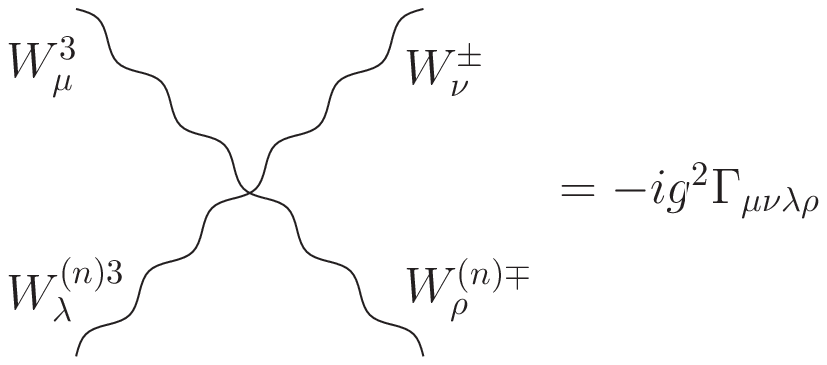} \ \ \ \
\includegraphics[width=3.5in]{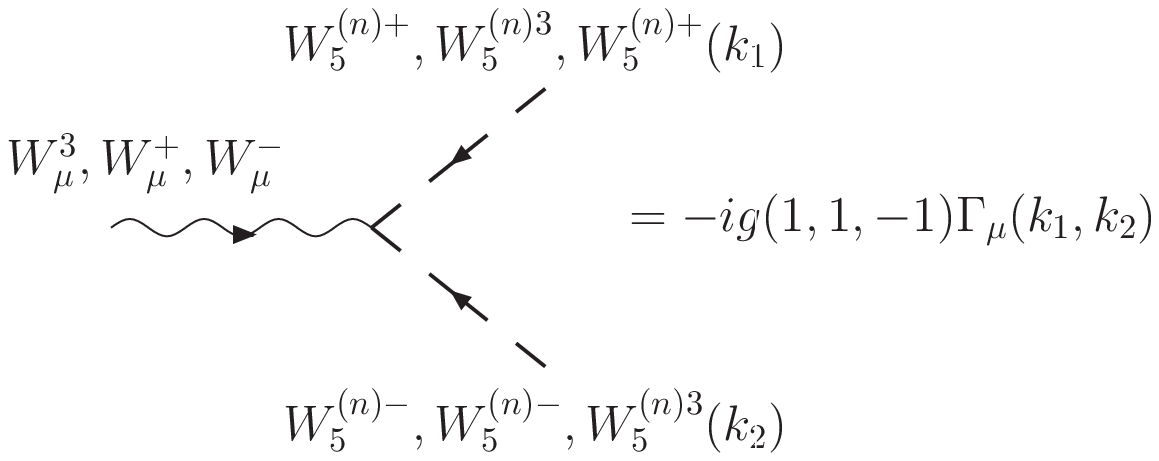}
\vskip 0.5cm
\includegraphics[width=2.5in]{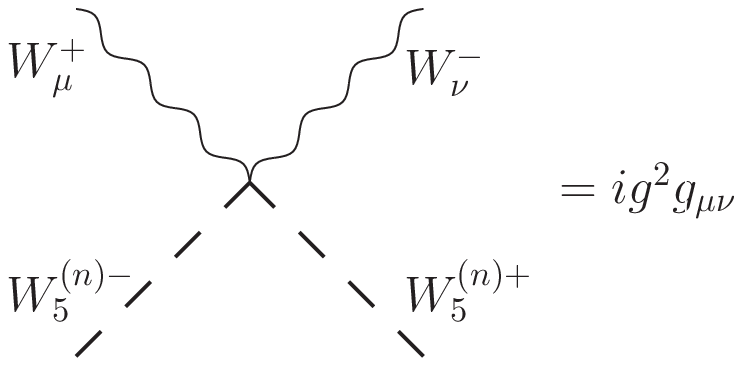} \ \ \ \
\includegraphics[width=2.5in]{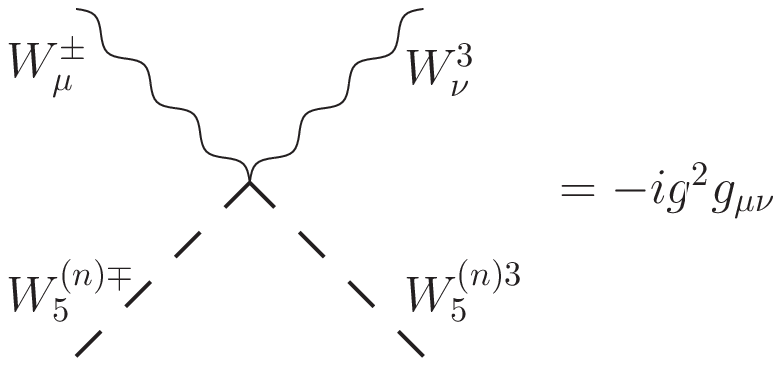}
\caption{\label{FR} Feynman rules for the vertices contributing to the $W^-W^+W^3$ coupling.}
\end{figure}

The Feynman rules needed to calculate the one-loop amplitude for the $W^-W^+W^3$ vertex can easily be derived from the above Lagrangians. The vertex functions that can contribute to the $W^-W^+W^3$ vertex at the one-loop level are shown in Fig.\ref{FR}. The couplings of zero KK gauge modes with excited modes of ghost-antighost fields are not shown, but they are identical to those with pseudo Goldstone bosons, which arises as a consequence of the invariance of the ghost sector under SGT. Also, due to such gauge invariance of the theory, the Lorentz tensor structure of the vertices characterizing the couplings $W^3W^{(n)-}W^{(n)+}$, $W^+W^{(n)-}W^{(n)3}$, and $W^-W^{(n)+}W^{(n)3}$ is the same. The vertex functions appearing in Fig.\ref{FR} are given by
\begin{equation}
  \Gamma_{\mu\lambda\rho}(k,k_1,k_2) = -(k_1-k_2)_\mu g_{\lambda\rho}+\bigg(k-k_2-\frac{1}{\xi}k_1\bigg)_\lambda g_{\rho\mu} -\bigg(k-k_1-\frac{1}{\xi}k_2\bigg)_\rho g_{\lambda\mu} \ ,
\end{equation}
\begin{equation}\label{}
    \Gamma_{\mu\nu\lambda\rho}=2g_{\lambda\mu}g_{\rho\nu}-g_{\mu\nu}g_{\lambda\rho}
    -\bigg(1+\frac{1}{\xi}\bigg)g_{\rho\mu}g_{\lambda\nu} \ ,
\end{equation}
\begin{equation}\label{}
    \Gamma_\mu(k_1,k_2)=(k_1-k_2)_\mu \ .
\end{equation}
Notice that as a consequence of gauge invariance, the trilinear gauge vertex satisfies the simple Ward identity
\begin{equation}
k^\mu \Gamma_{\mu \lambda \rho }(k,k_1,k_2)=\Gamma_{\lambda \rho}(k_1)-\Gamma_{\lambda \rho}(k_2)\, ,
\end{equation}
where
\begin{equation}
\Gamma_{\lambda \rho}(k)=(k^2-m^2_n)g_{\lambda \rho}-\left(1-\frac{1}{\xi}\right)k_\lambda k_\rho\,.
\end{equation}

\subsection{One-loop form factors of the $W^-W^+W^3$ vertex}
\begin{figure}
\centering\includegraphics[width=2.0in]{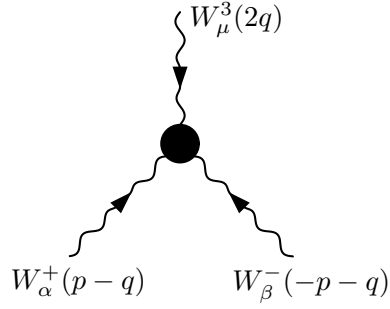}
\caption{\label{TLV} The trilinear $W^-W^+W^3$ vertex. The large black disc denotes loop contributions.}
\end{figure}

The vertex function for the $W^-W^+W^3$ coupling, with $W^3$ off-shell, can be parametrized in terms of three form factors as follows
\begin{eqnarray}
\Gamma^{W^3}_{\alpha \beta \mu}&=&-ig_{W^3}\Bigg\{ A\left[2p_\mu g_{\alpha \beta}+4\left(q_\beta g_{\alpha \mu}-q_\alpha g_{\beta \mu}\right)\right]\nonumber \\
&&+2\Delta \kappa_{W^3} \left(q_\beta g_{\alpha \mu}-q_\alpha g_{\beta \mu}\right)+\frac{4\Delta Q_{W^3}}{m^2_W}\left(p_\mu q_\alpha q_\beta -\frac{1}{2}q^2p_\mu g_{\alpha \beta}\right)\Bigg\}\, ,
\end{eqnarray}
where $g_{W^3}=gs_W$ for $W^3=\gamma$ and $g_{W^3}=gc_W$ for $W^3=Z$. We have introduced the short-hand notation $s_W$ and $c_W$ for the sine and cosine of the weak angle, respectively. We have dropped the CP-odd terms since they do not arise at the one-loop level in the context of the theory that we are considering. Our notation and conventions are shown in Fig. \ref{TLV}. In the SM, the tree--level values are $A=1$, $\Delta \kappa_{W^3}=0$, and $\Delta Q_{W^3}=0$. Within the context of a renormalizable theory, it is expected that the one-loop radiative corrections give divergent contributions to $A$, but finite ones to $\Delta \kappa_{W^3}$ and $\Delta Q_{W^3}$. The $A$ form factor is associated with the interaction
\begin{equation}
{\cal L}_{WWW^3}=-ig\left[\left(W^-_{\mu \nu}W^+\nu-W^+_{\mu \nu}W^{-\nu} \right)W^{3\mu}-W^3_{\mu \nu}W^{-\mu}W^{+\nu}\right]\, ,
\end{equation}
which is already present at the level of the classical action and must be, therefore, renormalized. Although the $\Delta \kappa_{W^3}$ form factor is associated with the interaction $ W^3_{\mu \nu}W^{-\mu}W^{+\nu}$, which is renormalizable, it arises as an anomalous contribution to the magnetic dipole and electric quadrupole moments of the $W^\pm$ gauge boson, similar to the one encountered for the case of spin $1/2$ charged particles. This contribution is always finite in a renormalizable theory. The form factor $\Delta Q_{W^3}$ is associated with the dimension-six interaction  $W^{-\rho}_\lambda W^+_{\rho \sigma}W^{3\lambda \sigma}$, which necessarily arises at one-loop in a renormalizable theory, being therefore a finite prediction of the theory under consideration. Although theories in more than four dimensions are nonrenormalizable, we will show below that the one-loop contributions of the KK modes to these form factors resemble those of a renormalizable theory.

\begin{figure}
\centering\includegraphics[width=1.5in]{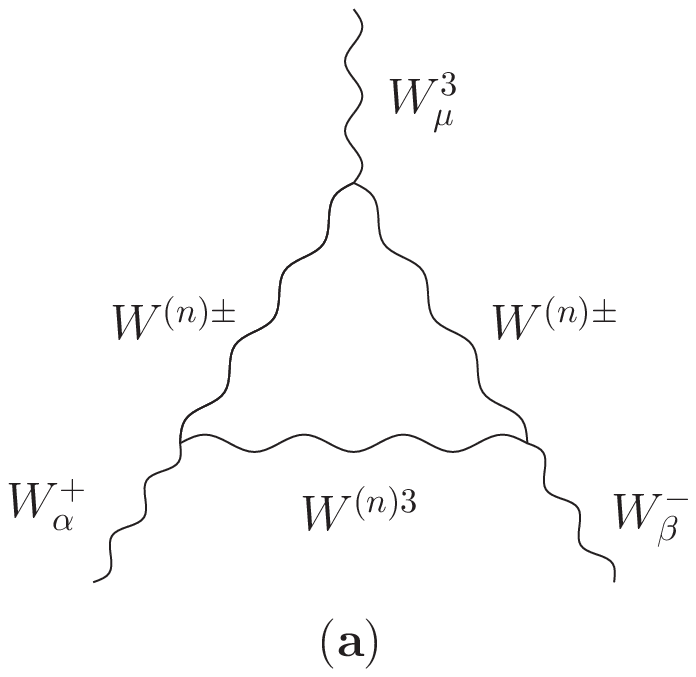}
\centering\includegraphics[width=1.3in]{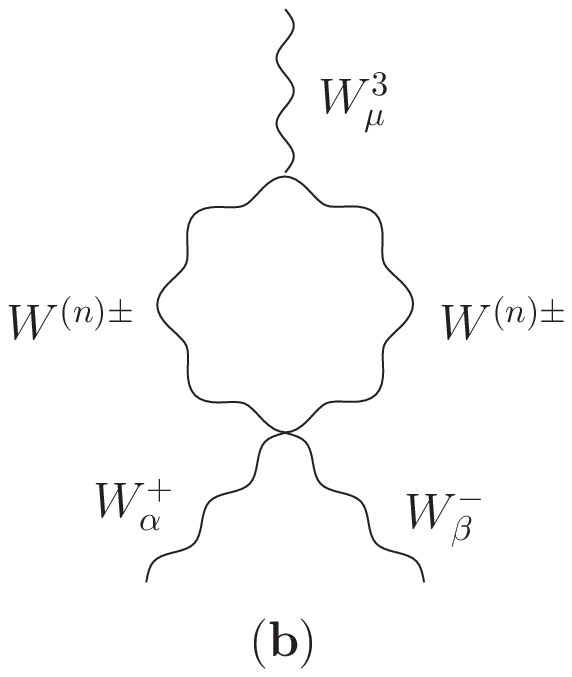}
\centering\includegraphics[width=1.5in]{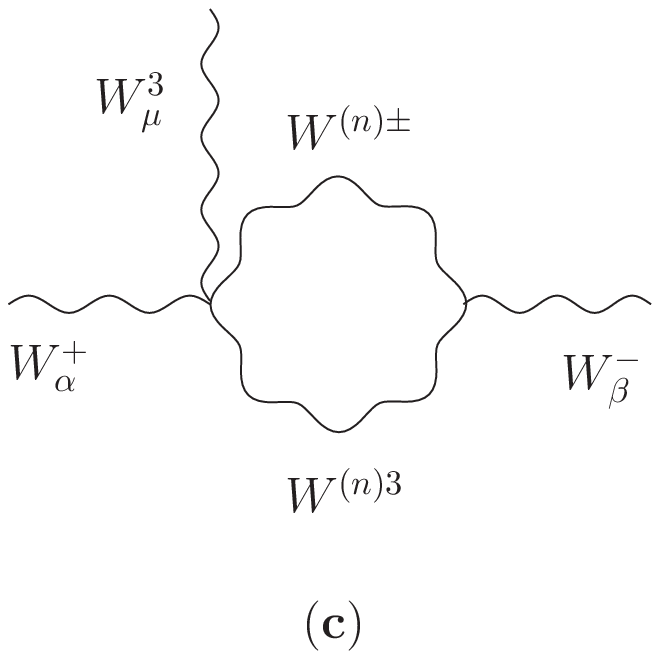}
\centering\includegraphics[width=1.5in]{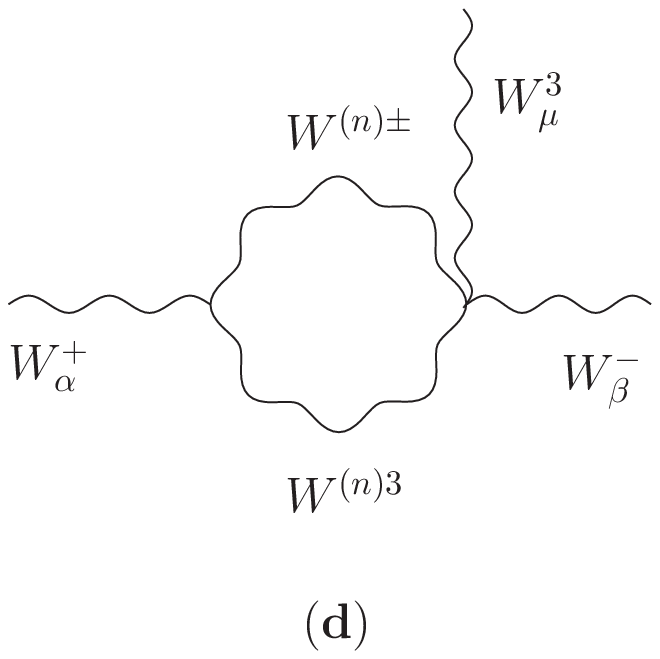}
\caption{\label{FD}Feynman diagrams contributing to the $W^-W^+W^3$ coupling. Identical diagrams for the contributions of pseudo Goldstone bosons and ghost fields have not been displayed.}
\end{figure}

The Feynman diagrams that can contribute to the $W^-W^+W^3$ Green function are shown in Fig. \ref{FD}. As it has been emphasized, our treatment of this coupling is invariant under the SGT of $SU_L(2)$, although gauge dependent, which also occurs, for example, when the BFM is applied. In other words, gauge-invariant quantum actions render gauge-invariant but not gauge-independent Green functions. However, motivated by the link between the BFM and the PT, as well as the previous analysis presented in Refs.~\cite{CG331}, we will present our results in the Feynman-'t Hooft gauge. The calculation in this class of nonlinear gauge is simple indeed. In particular, we found that this coupling only receives contributions from triangle diagrams, as the bubble diagrams contributions are identically zero each one. In addition, in this nonlinear gauge, the contribution of the ghost-antighost sector is exactly minus twice the pseudo Goldstone boson contribution. Thus, the net contribution is equal to the contribution of the excited $W^{(n)}$ modes minus the pseudo-Goldstone boson one. Such a contribution can be written as follows
\begin{eqnarray}
 A &=& \frac{g^2}{96\pi^2(4x_W-1)^3} \sum_{n=1}
        \bigg(
        -24 x_n (1-4 x_W)^2 B_0(1)
        +48 x_W [5 (x_W-1) x_W \nonumber \\
    &&  +4 x_n (4 x_W-1)] B_0(2)
        -6 [-4 x_W (6 x_W^2-2 x_W+3)+x_n (64 x_W^2-4)+1] B_0(3) \nonumber \\
    &&  +36 Q^2 x_W [-4 x_W^3+2 x_W^2+x_W+4 x_n (x_W-1) (4 x_W-1)] C_0 \nonumber \\
    &&  -4 \{1-2 x_W [x_W (20 x_W-33)+9]\}
        \bigg) ,
\end{eqnarray}

\begin{eqnarray}
\Delta\kappa_{W^3} &=& \frac{g^2}{96\pi^2(4x_W-1)^3} \sum_{n=1}
                 \bigg\{
                 48 x_n [B_0(1)-B_0(2)]
                 +12 x_W (26 x_W+1) [B_0(2)-B_0(3)] \nonumber \\
             &&  -384 x_n x_W [B_0(1)-2 B_0(2)+B_0(3)]
                 +768 x_n x_W^2 [B_0(1)-3 B_0(2)+2 B_0(3)] \nonumber\\
             &&  -72 Q^2 x_W (x_W^2-12 x_n x_W+x_W+3 x_n) C_0
                 +12 x_W (4 x_W-1) (8 x_W+3)
                 \bigg\} ,
\end{eqnarray}

\begin{eqnarray}
\Delta Q_{W^3} &=& \frac{g^2}{96\pi^2(4x_W-1)^3} \sum_{n=1}
            \bigg\{
            -1536 x_n x_W^3 [B_0(1)-B_0(3)]
            +96 (x_W-1) x_W^2 (6 x_W+1) [B_0(2)-B_0(3)] \nonumber\\
         && +768 x_n x_W^2 [B_0(1)+B_0(2)-2 B_0(3)]
            -96 x_n x_W [B_0(1)+2 B_0(2)-3 B_0(3)] \nonumber\\
         && +144 Q^2 x_W [4 x_W^4-4 (4 x_n+1) x_W^3+2 (6 x_n+1) x_W^2-6 x_n x_W+x_n] C_0 \nonumber\\
         && -24 x_W (4 x_W-1) [2 x_W (6 x_W-1)+1]
             \bigg\} ,
\end{eqnarray}
where we have introduced the definitions $Q\equiv 2q$, $x_W\equiv m_W^2/Q^2$, and $x_n\equiv m_n^2/Q^2$, with $m_n=n/R$, $n=1,2,...$. In addition, we have introduced the following short-hand notation for the Passarino-Veltman scalar functions: $B_0(1)\equiv B_0(0,m_n^2,m_n^2)$, $B_0(2)\equiv B_0(m_W^2,m_n^2,m_n^2)$, $B_0(3)\equiv B_0(Q^2,m_n^2,m_n^2)$, and $C_0\equiv C_0(m_W^2,m_W^2,Q^2,m_n^2,m_n^2,m_n^2)$. Notice that the form factor $A$ is divergent, but the $\Delta \kappa_{W^3}$ and $\Delta Q_{W^3}$ ones are free of ultraviolet divergences. As far as the discrete sum is concerned, we will show in the next section that it is convergent for the case of the $\Delta \kappa_{W^3}$ and $\Delta Q_{W^3}$ form factors. The same is true for the divergent form factor $A$. In this case, effects of heavy KK modes disappear, as they can be absorbed by the parameters of the light theory~\cite{Ren}. It is worth mentioning the fact that the excited KK modes together with their associated pseudo Goldstone bosons and ghosts separately lead to finite and gauge invariant results, which is a peculiarity of this class of gauge-fixing procedures~\cite{CG331}. To conclude this part, let us comment an important aspect concerning gauge invariance. It has been shown above that the amplitude for the $W^-W^+W^3$ coupling can be written as
\begin{equation}
\Gamma^{W^3}_{\alpha \beta \mu}=-g_{W^3}I_{\alpha \beta \mu} \, ,
\end{equation}
with $I_{\alpha \beta \mu}$ the loop amplitude, which is the same for both the $WW\gamma$ and the $WWZ$ couplings. The important point to be emphasized here is the fact that the Green functions for these vertices differ only by the factor $g_{W^3}$, just as it occurs at the level of the classical action, which means that the $SU_L(2)$ symmetry is preserved at the one-loop level. For on-shell $W$ bosons, this Green function satisfies the Ward identity
\begin{equation}
q^\mu \Gamma^{W^3}_{\alpha \beta \mu}=0\, .
\end{equation}

\section{Discussion}
\label{D} Before presenting our numerical results, let us to comment the importance of the $W^-W^+\gamma$ and $W^-W^+Z$ vertices in collider physics. Within this respect,  diverse studies~\cite{LHC-ILC} point out that the sensitivity of LHC is of $O(10^{-2})$ for $\Delta \kappa_{W^3}$ and of up to $O(10^{-3})$ for $\Delta Q_{W^3}$. A higher sensitivity is expected at the ILC. The most stringent bounds are obtained from angular distributions in $W$ pair production, which lead to bounds of $O(10^{-4})$ for both $\Delta \kappa_{W^3}$ and $\Delta Q_{W^3}$~\cite{LHC-ILC}. Although the LHC will be capable of producing new particles in a real way due its high center-of-mass energy, the ILC machine offers important advantages for precision studies. An important advantage of this collider is the fact that the energy of the colliding beams is exactly known. Variable tuning of the beam energy together with the control over its polarization would provide valuable information concerning the presence of new physics. Due to this, precision studies on the trilinear $W^-W^+W^3$ vertex that eventually reveal the presence of new particles will be more accessible to ILC. In this context, the reaction $e^+e^-\to W^+W^-$, with the $Z$ and the photon gauge bosons highly off-shell, will play a decisive role in future researches at this collider, as it will provide relevant information for our knowledge of the SM such as a more precise determination of the $W$ mass and its width decays, and it will also open up the possibility for detecting new physics effects via the distinctive s-channel contribution from the $W^-W^+W^3$ vertex. Radiative corrections to the $e^+e^-\to W^+W^-$ processes have been calculated in the context of the SM both for on-shell~\cite{SMONS} and off-shell~\cite{SMOFFS} $W$ gauge bosons. Beyond the SM, this reaction has been studied in the context of supersymmetric theories~\cite{SUSY}, technicolor models~\cite{TCM}, and also in a model-independent way using the effective Lagrangian approach~\cite{ELA}.

As already mentioned in the introduction, the radiative corrections to the $W^-W^+W^3$ vertex with the two $W$ gauge bosons on-shell and the $W^3$ one off-shell, have received considerable attention from both the phenomenological and theoretical points of view. In the SM, the one--loop amplitudes were calculated using the conventional linear gauge along with the Feynman--'t Hooft gauge~\cite{TLVSM}. As emphasized in that work, the resultant
amplitudes are not gauge-invariant, which is evident from the presence of infrared divergences and the bad high-energy behavior of the $\Delta \kappa_{W^3}$ form factor. In contrast, it was found that $\Delta Q_{W^3}$ is well-behaved~\cite{TLVSM}. Subsequently, these vertices were revisited by Papavassiliou and Philippides~\cite{PTWWV1} in a gauge-invariant way via the PT, finding that the form factor $\Delta \kappa_{W^3}$ disagrees from that presented in~\cite{TLVSM}, though there is agreement for $\Delta Q_{W^3}$. It was found that for energies in the range $200 \, \mathrm{GeV} <Q<1000\, \mathrm{GeV}$, $\Delta \kappa_{\gamma}$ goes from $10^{-3}$ to $10^{-4}$, whereas $\Delta Q_{\gamma}$ ranges from $10^{-4}$ to $6\times 10^{-5}$ in the same range of energies~\cite{TLVSM,PTWWV1}. Although the presence of new physics may improve these values, it is not the case however of supersymmetry, which predicts similar or smaller contributions than the SM ones~\cite{TLVSUSY}. More recently, the radiative corrections to these vertices due to new gauge bosons were studied in Ref.~\cite{CG331}, within the context of the so-called 331 models, using a $SU_L(2)$-covariant gauge-fixing procedure which, as we already emphasized, is quite similar to the one used in this work. Well-behaved amplitudes were encountered at high energies. In particular, the new physics effects decouple in the large mass limit. It was found that for relatively light new gauge bosons, with masses in the range $2m_W<m_Y<8m_W$ and energies varying in the domain $200 \, \mathrm{GeV} <Q<1000\, \mathrm{GeV}$, both $\Delta \kappa_{W^3}$ and $\Delta Q_{W^3}$ go from $10^{-4}$ to $10^{-5}$. These results are of the same order of magnitude than those predicted by the SM. These form factors are of the order of $10^{-6}$ for the same range of energies and heavier gauge bosons~\cite{CG331}.

\begin{figure}
\centering
\includegraphics[width=3.0in]{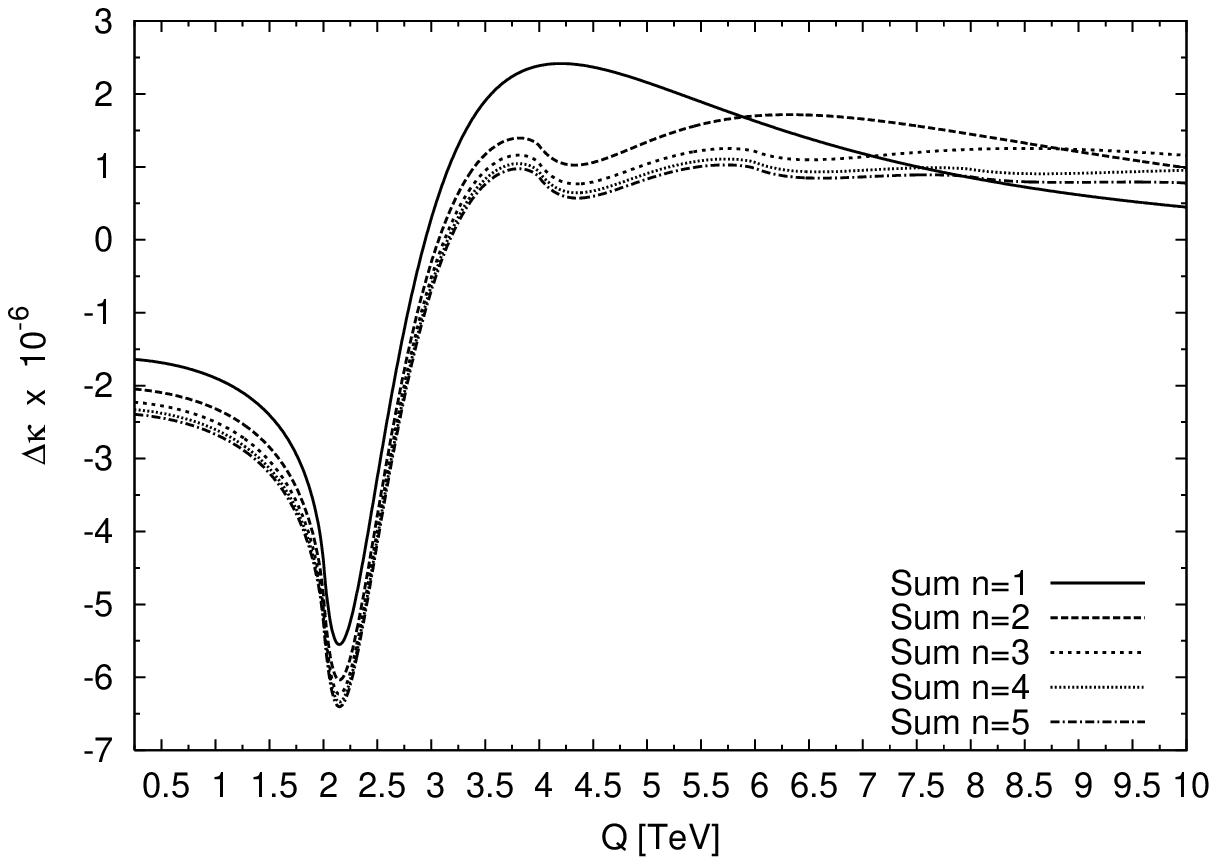}
\includegraphics[width=3.0in]{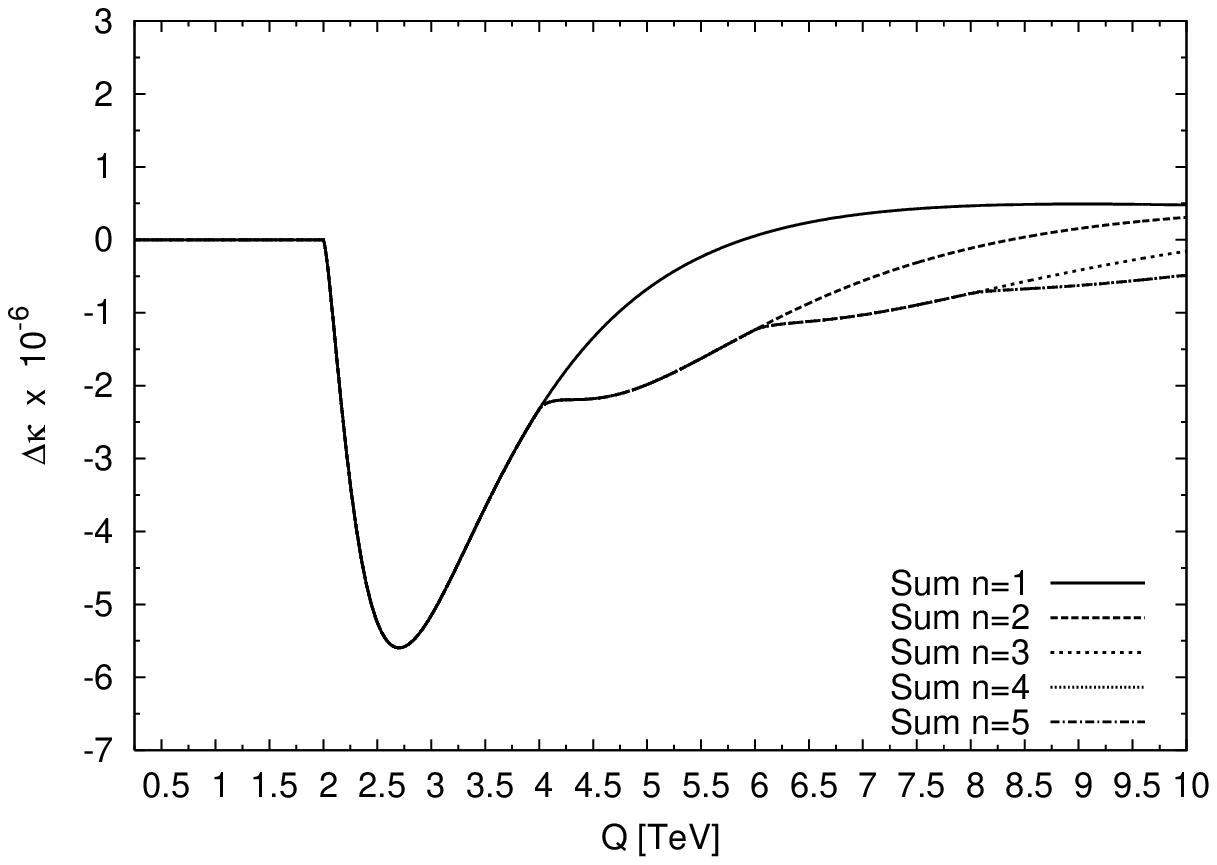}
\caption{\label{gd1} The behavior of real (left graphic) and imaginary (right graphic) parts of $\Delta \kappa$ as a function of the energy for $R^{-1}=1$ TeV. The contributions of up to five excited KK modes were considered.}
\end{figure}

\begin{figure}
\centering
\includegraphics[width=3.0in]{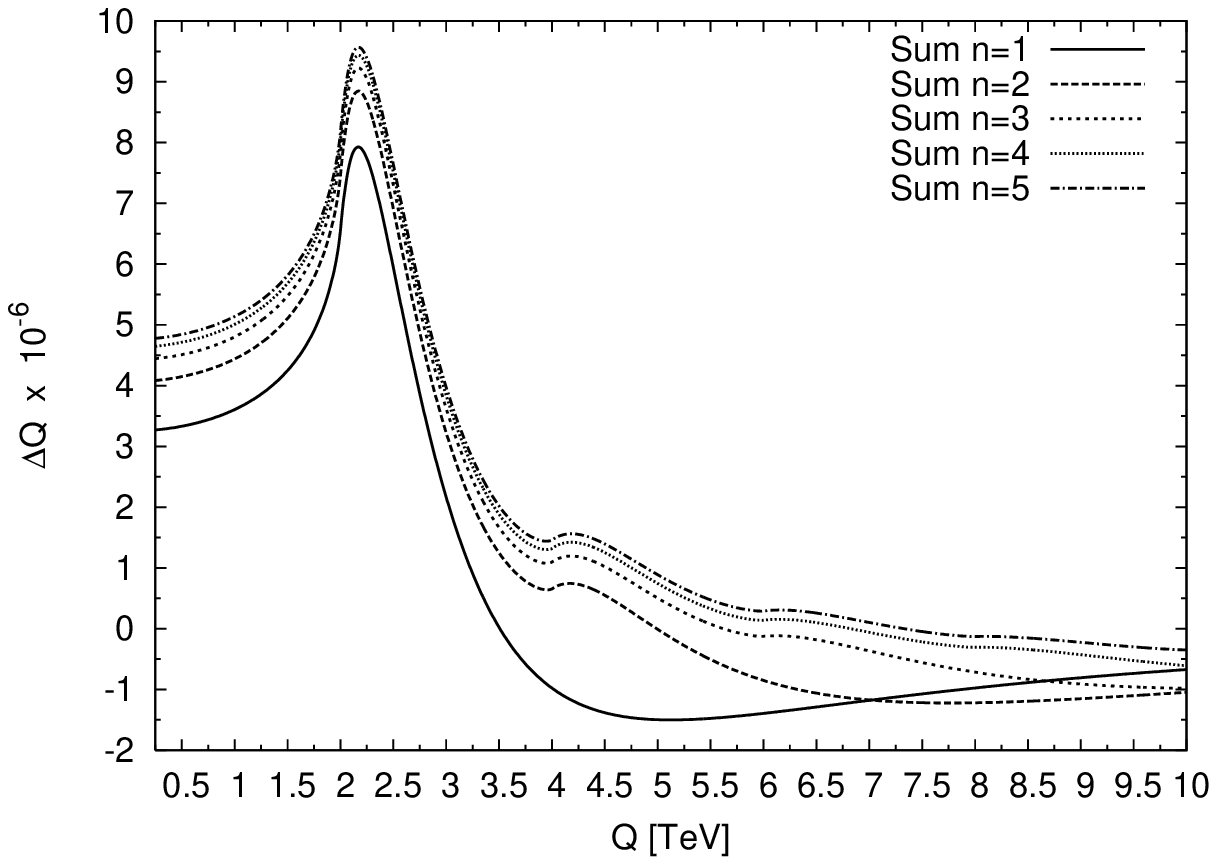}
\includegraphics[width=3.0in]{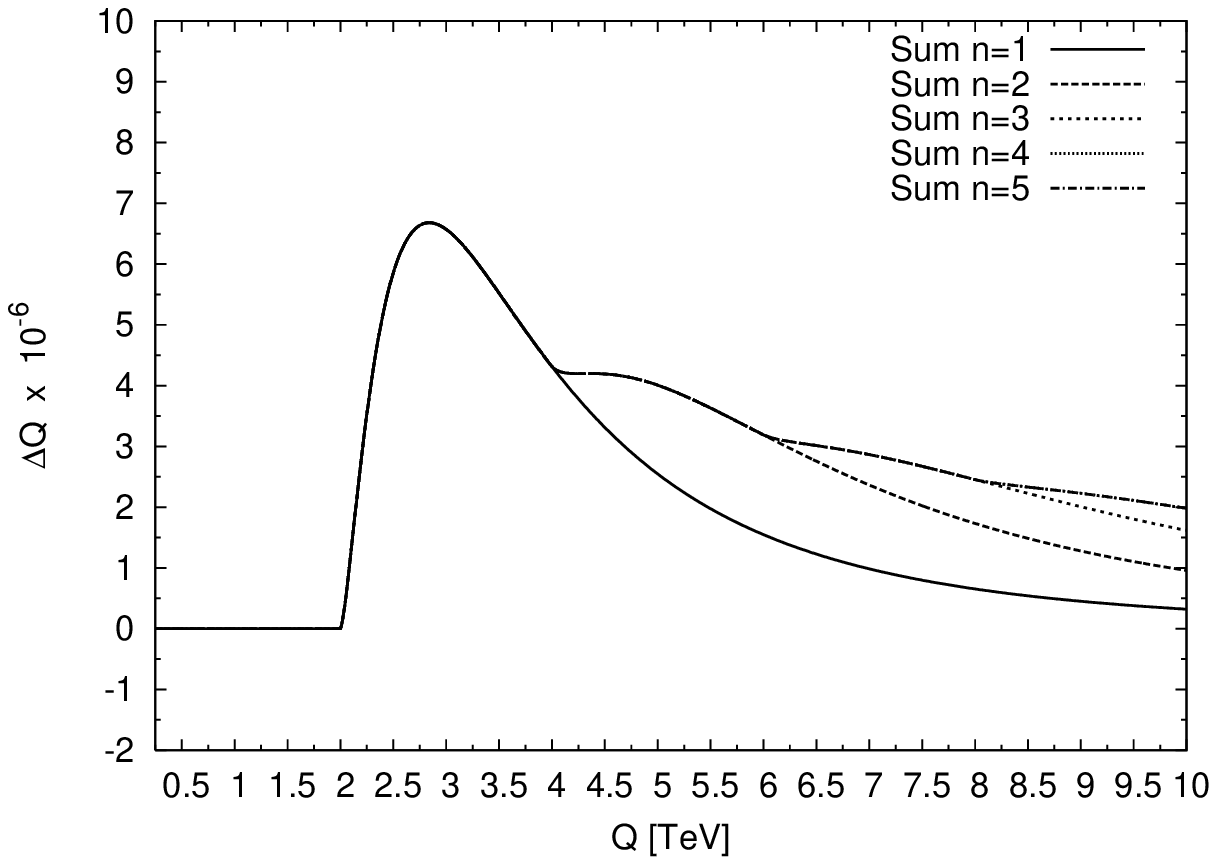}
\caption{\label{gq1} The same than in Fig. \ref{gd1} but now for $\Delta Q$.}
\end{figure}

\begin{figure}
\centering
\includegraphics[width=3.0in]{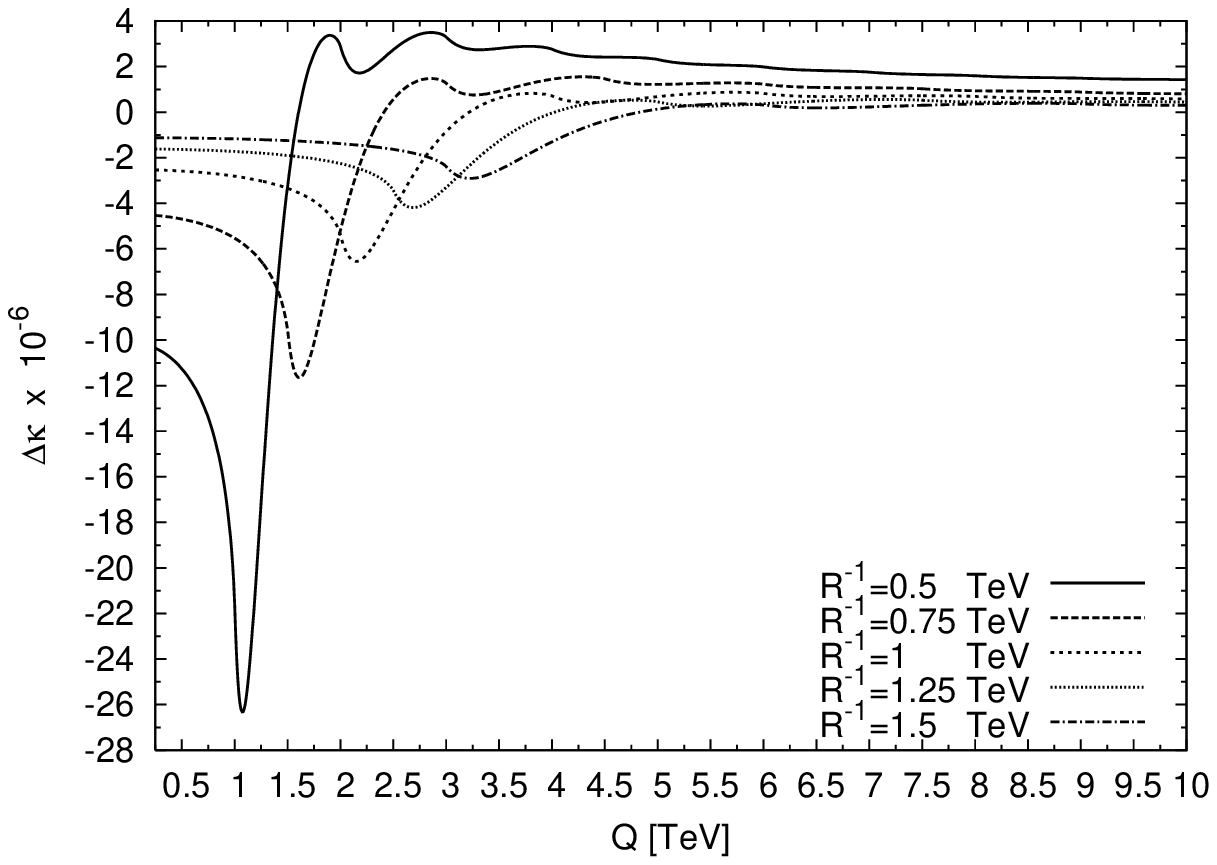}
\includegraphics[width=3.0in]{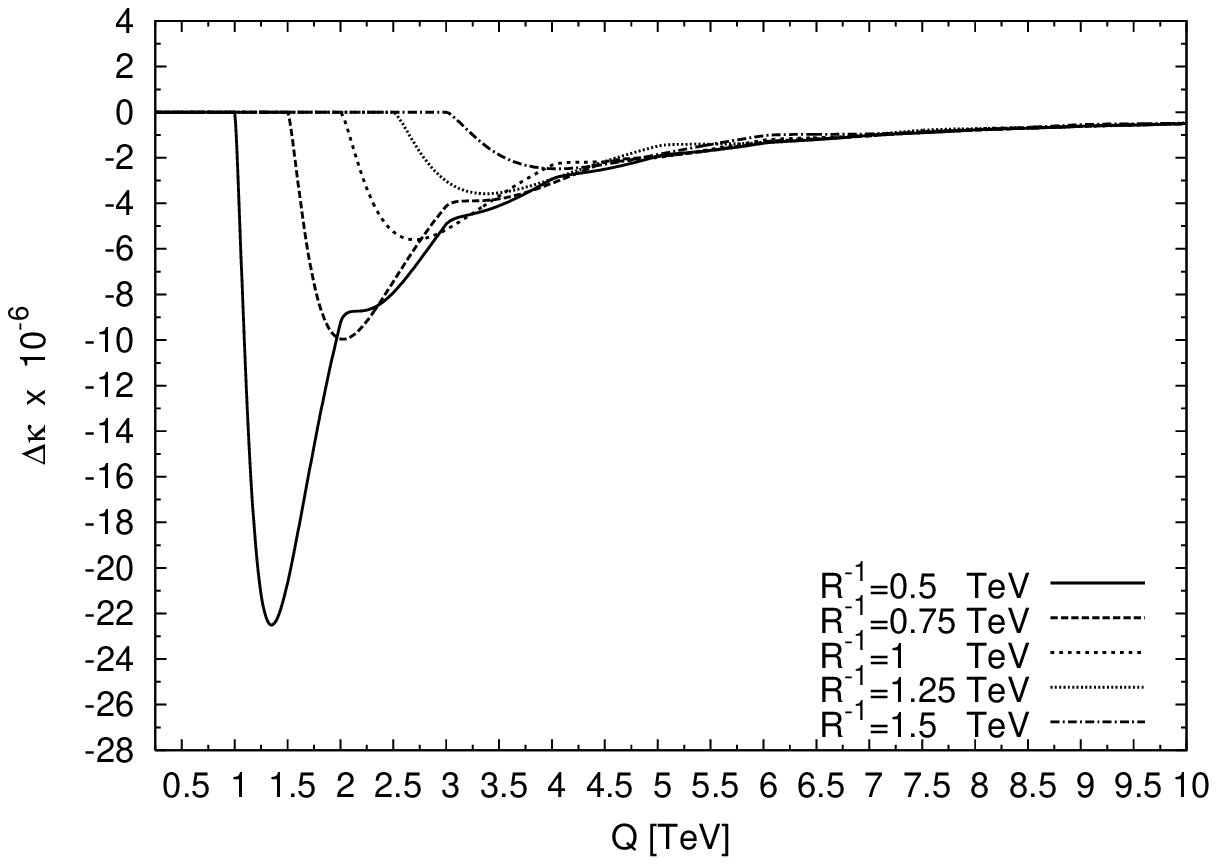}
\caption{\label{DR} The same than in Fig. \ref{gd1} but now for several values of $R^{-1}$. The sum of the first 10 excited KK modes was considered.}
\end{figure}

\begin{figure}
\centering
\includegraphics[width=3.0in]{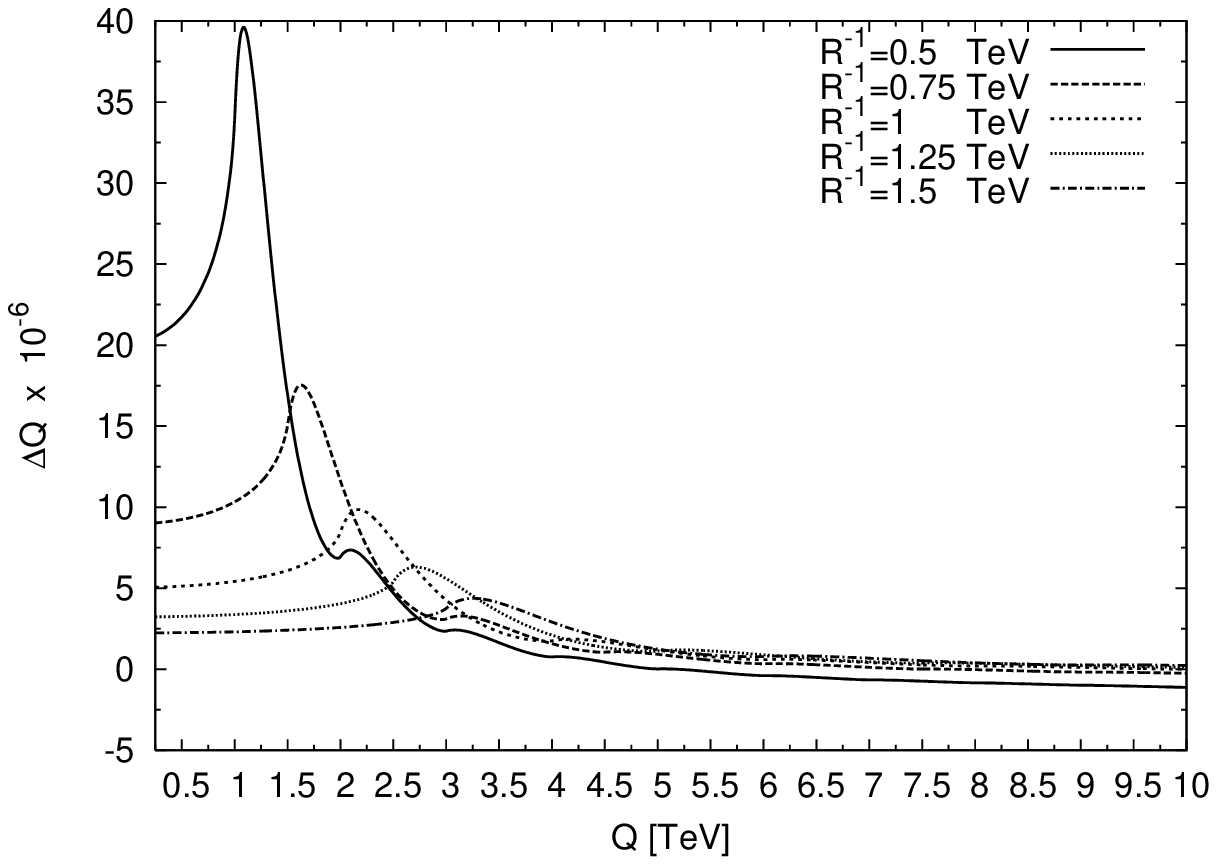}
\includegraphics[width=3.0in]{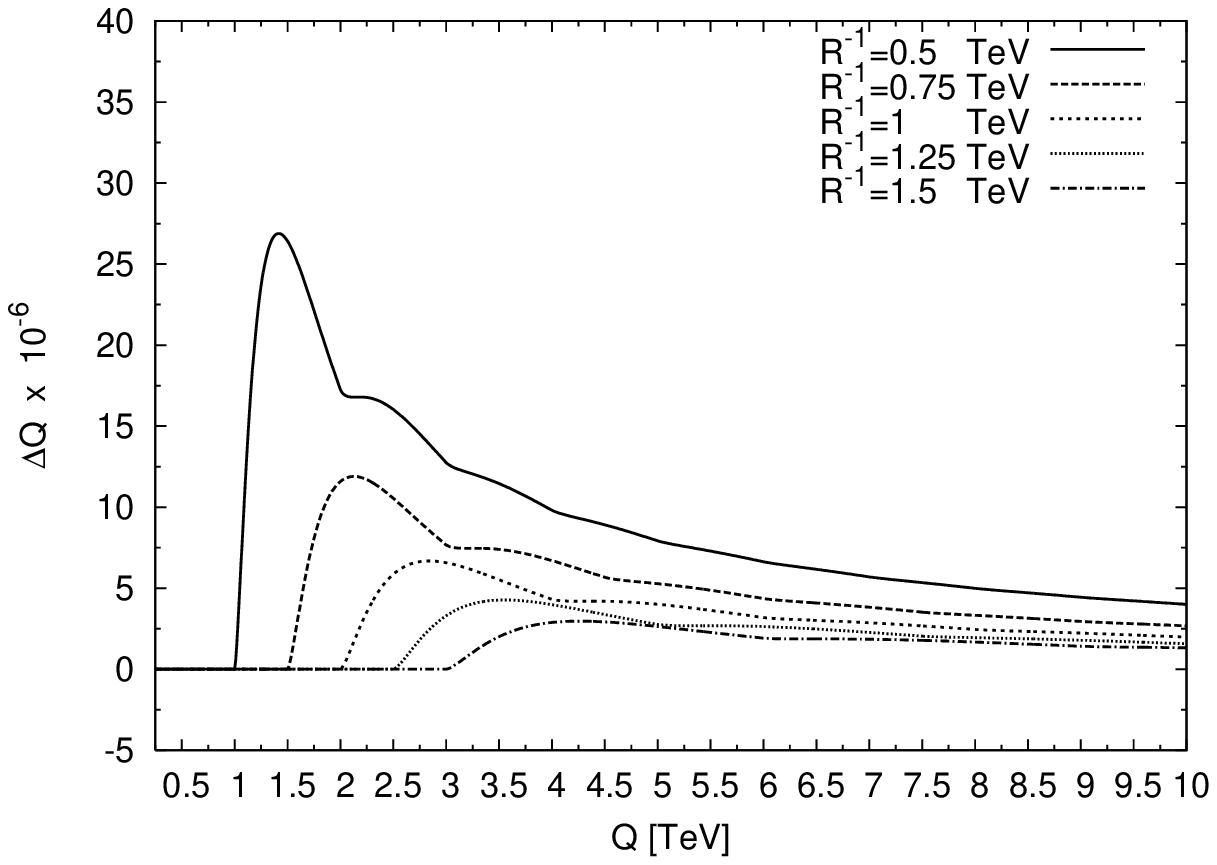}
\caption{\label{QR} The same than in Fig. \ref{DR} but now for $\Delta Q$.}
\end{figure}

\begin{figure}
\centering
\includegraphics[width=3.0in]{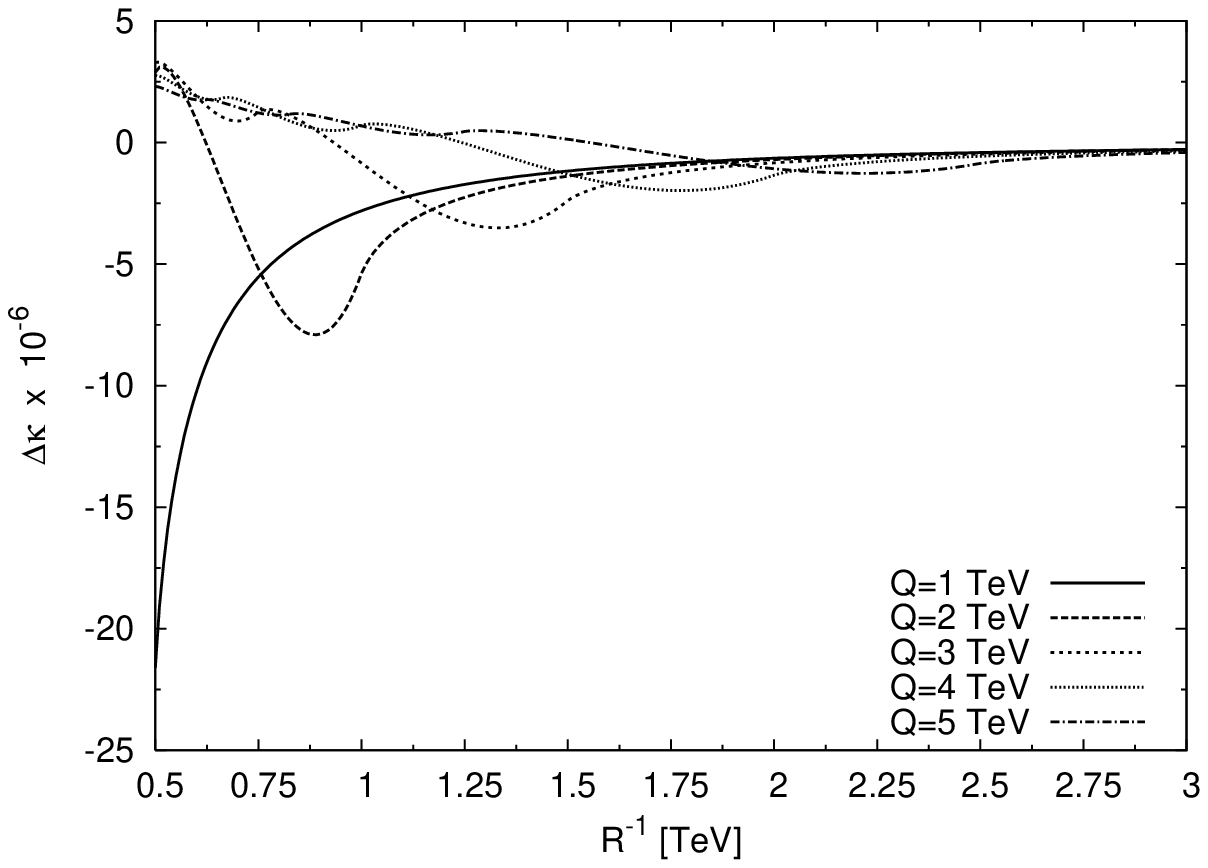}
\includegraphics[width=3.0in]{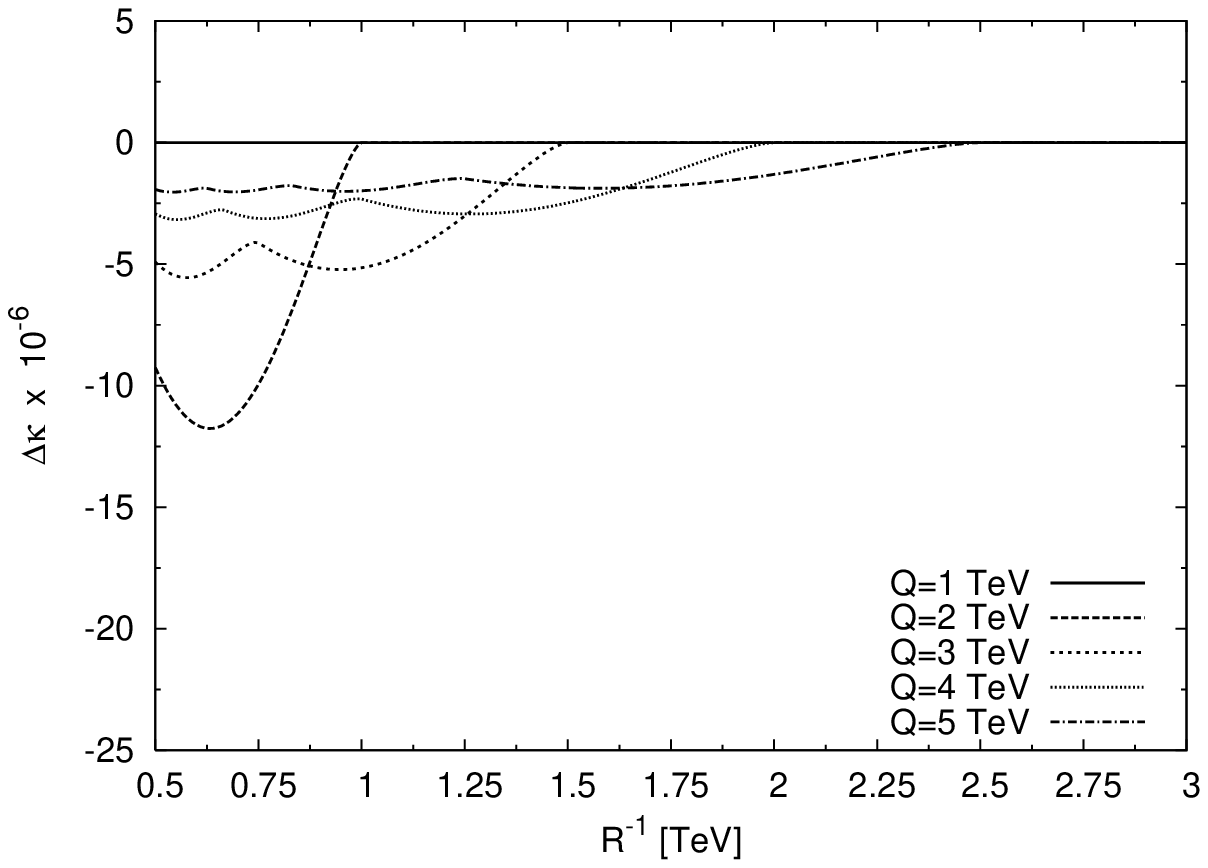}
\caption{\label{DD} Behavior of the real (left graphic) and imaginary (right graphic) parts of $\Delta \kappa$ as a function of $R^{-1}$ for some values of the energy $Q$. The contribution of the first 10 excited KK modes was considered.}
\end{figure}

\begin{figure}
\centering
\includegraphics[width=3.0in]{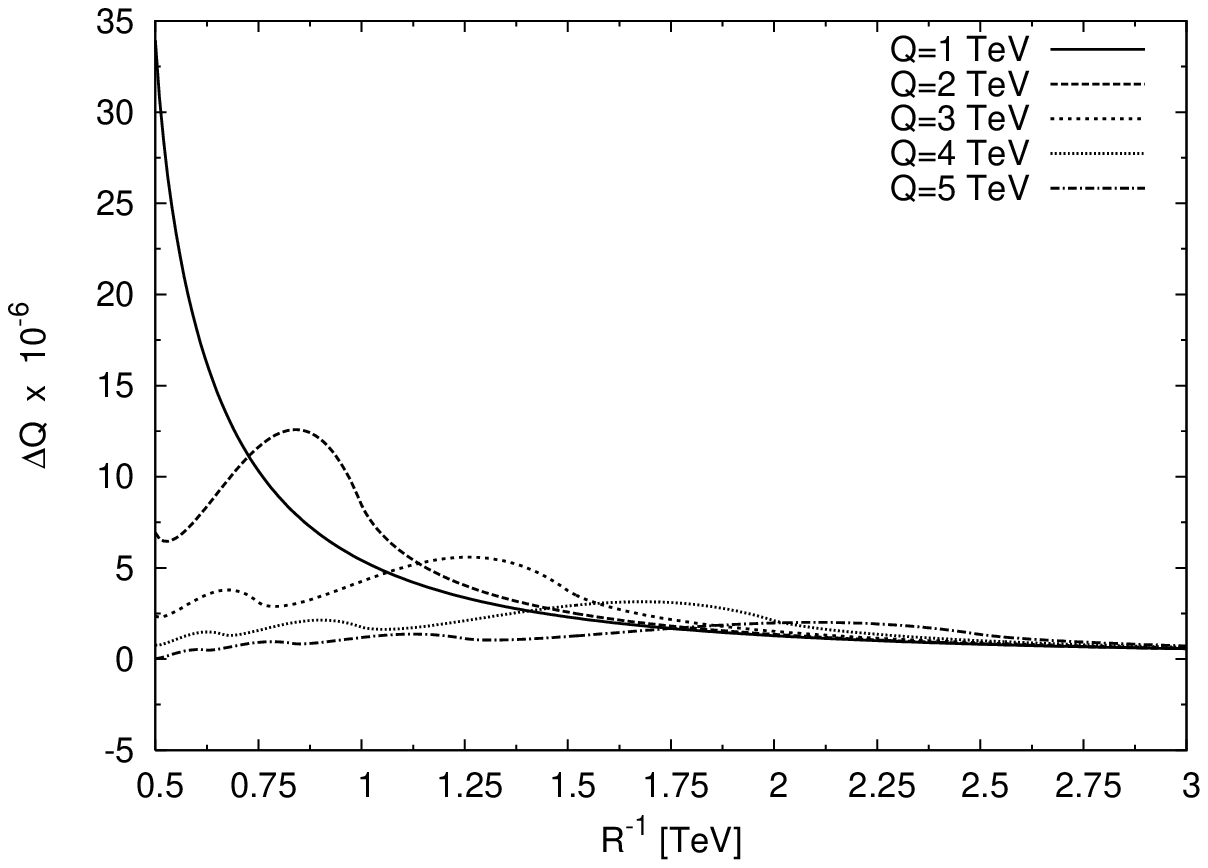}
\includegraphics[width=3.0in]{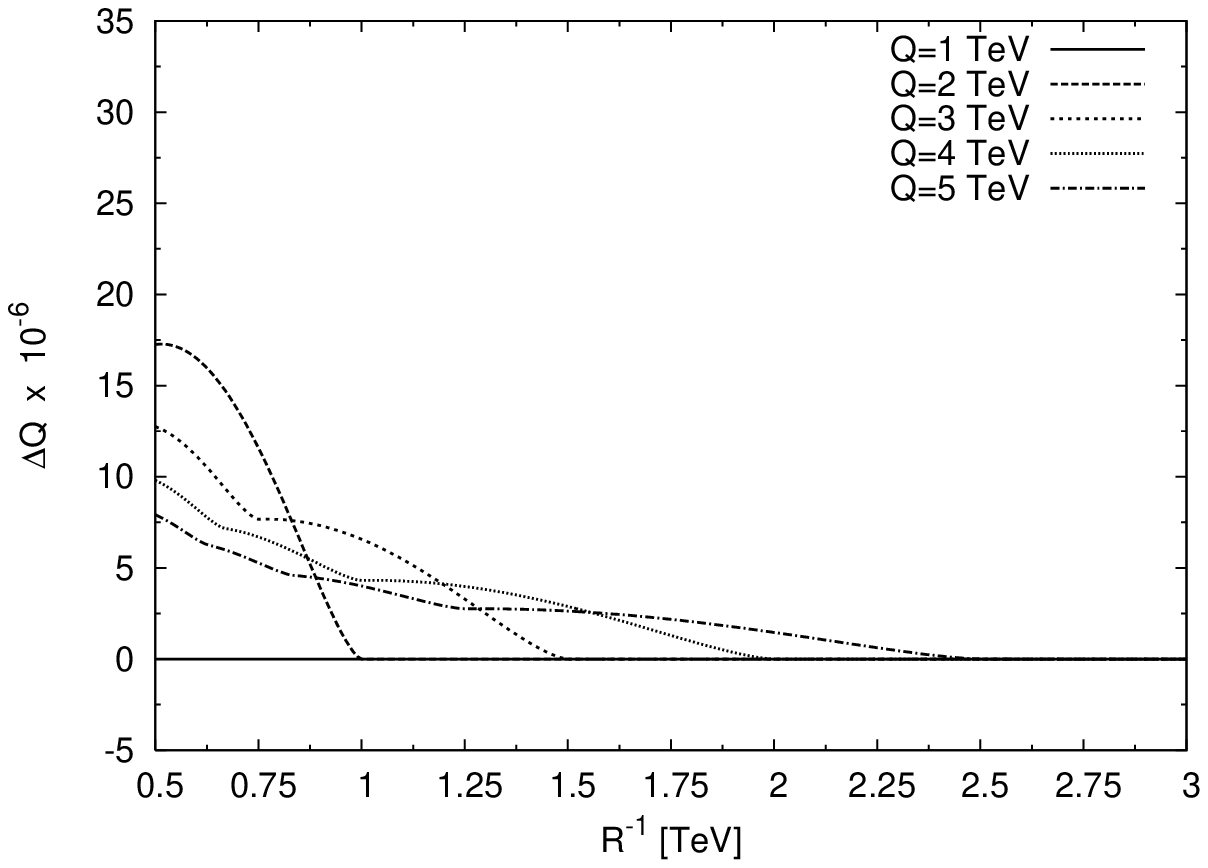}
\caption{\label{DQ} The same than in Fig. \ref{DD} but now for $\Delta Q$.}
\end{figure}

We now turn to present and discuss our numerical results. In Fig. \ref{gd1}, the behavior of the real (left graphic) and imaginary (right graphic) parts of the $\Delta \kappa$\footnote{From now on, the form factors $\Delta \kappa_{W^3}$ and $\Delta Q_{W^3}$ will be denoted simply by $\Delta \kappa$ and $\Delta Q$, respectively.} form factor are shown as functions of the energy $Q$. The same is shown in Fig. \ref{gq1} but for the $\Delta Q$ form factor. The value of $R^{-1}=1$ TeV for the compactification scale has been taken. The sum of the KK modes contributions has been considered. In these figures five cases have been taken into account, namely the contribution of only one KK mode $n=1$, the sum of the contributions of the first two excited modes $n=2$, the sum of the contributions of the first three excited modes $n=3$, etc. For this value of the new physics scale, the contributions of excited modes beyond $n=5$ are indeed insignificant. It is interesting to study the behavior of these form factors for several values of the scale $R$. In Figs. \ref{DR} and \ref{QR}, the behavior of the real and imaginary parts of $\Delta \kappa$ and $\Delta Q$ are shown for some values of the scale $R$, where the contribution of the first 10 excited KK modes has been taken into account. From these figures, it can be appreciated that for $0.5\, \mathrm{TeV}<R^{-1}<1\, \mathrm{TeV}$,  both $\Delta \kappa$ and $\Delta Q$ range from approximately $10^{-5}$ to $10^{-6}$, at the best. These results are of the same order of magnitude than those arising from the new gauge bosons appearing in the 331 model~\cite{CG331}, which however should be compared with those obtained from the one-loop radiative correction in the context of the SM, which are $\Delta \kappa\approx 4\times 10^{-3}$ and $\Delta Q\approx 4\times 10^{-5}$~\cite{CNg}. Finally, we would like to comment on the decoupling character of the KK modes contributions to light Green functions. In Figs. \ref{DD} and \ref{DQ}, the behavior of $\Delta \kappa$ and $\Delta Q$ as functions of the scale $R^{-1}$ is shown for some values of the energy $Q$. From these figures, it can be appreciated the decoupling nature of this type of new physics, as for fixed energy $Q$ these contributions quickly  disappear. We would like to stress that this well-behavior of the $\Delta \kappa$ and $\Delta Q$ form factors at high energies and their decoupling nature for large $R^{-1}$ show in essence that our quantum treatment of KK modes is correct and that it can be used to predict one-loop effects of extra dimensions on light Green functions.

\section{Effects of higher dimensional operators}
\label{hd} As commented in the introduction, gauge theories in more than four dimensions are not renormalizable under the power counting Dyson's criterion. They are not fundamental in this sense. Consequently, there is no criterion to limit the number of Lorentz and gauge invariants that can be present in the theory. In previous sections, the one-loop radiative correction of the KK modes to the $WWW^3$ vertex was studied. We now proceed to study the tree--level effects on this vertex induced by operators of canonical dimension higher than $5$. The lower dimensional operator that can contribute to both $\Delta \kappa_{W^3}$ and $\lambda_{W^3}$ form factors at the tree level has canonical dimension $15/2$, which is given by
\begin{equation}
\frac{g_5\alpha_W}{M^2_s}\frac{\epsilon^{abc}}{3!}{\cal W}^a_{\lambda \rho}{\cal W}^{b\rho\nu}{\cal W}^{c\  \lambda}_\nu\, ,
\end{equation}
where $3!$ is a symmetry factor and $\alpha_W$ is an unknown dimensionless parameter which depends on the details of the underlying physics. Notice that the presence of the  $g_5/M^2_s$ factor is needed to correct dimensions. This means that after integrating the compactified coordinate the effects induced by this operator will be suppressed by a factor of $M^2_s$ with respect to those induced by the dimension-five Yang-Mills theory studied previously.

Once integrated the fifth dimension and conserving up to terms involving only the zero mode, one obtains
\begin{equation}
\delta {\cal L}^{SU(2)}_{\rm YM}=\frac{g\alpha_W}{M^2_s} {\cal O}_{W}\, ,
\end{equation}
where ${\cal O}_W$ is a dimension-six operator given by
\begin{equation}
{\cal O}_W=\frac{\epsilon^{abc}}{3!}W^a_{\lambda \rho}W^{b\rho \nu}W^{c\ \lambda}_\nu\, .
\end{equation}
This operator is well-known in other contexts of new physics~\cite{ABHRT}. After some algebra, one obtains the following Lagrangian for the $W^-W^+W^3$ vertex:
\begin{equation}
{\cal L}^{\rm tree}_{WWW^3}=\frac{ig_{W^3}\lambda}{m^2_W}W^-_{\lambda \mu}W^{+\mu}_{\ \nu}W^{3\nu \lambda} \, ,
\end{equation}
where
\begin{equation}
\lambda=\alpha_W\Big(\frac{m_W}{M_s}\Big)^2\, .
\end{equation}
From the above Lagrangian, it is easy to derive the tree--level contributions to the $\Delta \kappa$ and $\Delta Q$ form factors, which are given by
\begin{eqnarray}
\Delta \kappa^{\rm tree}&=&- \lambda \, ,\\
\Delta Q^{\rm tree}&=&2\lambda\, .
\end{eqnarray}
For theories with only one extra dimension, $M_s$ is estimated to be of about $10^2R^{-1}$~\cite{Papa,Apel}, so $\lambda=10^{-4}\alpha_W (Rm_W)^2$. Assuming that $\alpha_W\sim O(1)$, we find that $|\Delta \kappa^{\rm tree}|$ ($\Delta Q^{\rm tree}$) ranges from $2.6\times 10^{-6}$ to $0.76\times 10^{-6}$ (from $5.2\times 10^{-6}$ to $1.3\times 10^{-6}$) for $0.5\ \mathrm{TeV}<R^{-1}<1.0\ \mathrm{TeV}$. It is interesting to compare these results with those induced at the one-loop level by the KK modes. In a scenario with $\sqrt{Q^2}=0.5\ \mathrm{TeV}$, $|\Delta \kappa^{\rm 1-loop}|$ ($|\Delta Q^{\rm 1-loop}|$) ranges from $1.12\times 10^{-5}$ to $2.58\times 10^{-6}$ (from $2.17\times 10^{-5}$ to $5.12\times 10^{-6}$ ) for $0.5 \ \mathrm{TeV}<R^{-1}<1.0\ \mathrm{TeV}$. In a scenario with $\sqrt{Q^2}=1.0\ \mathrm{TeV}$, there is a light variation, namely, $|\Delta \kappa^{\rm 1-loop}|$ ($|\Delta Q^{\rm 1-loop}|$) ranges from $2.16\times 10^{-5}$ to $2.81\times 10^{-6}$ (from $3.4\times 10^{-5}$ to $5.42\times 10^{-6}$ ) for $0.5 \ \mathrm{TeV}<R^{-1}<1.0\ \mathrm{TeV}$. This shows that the one-loop effect of the KK modes is about one order of magnitude larger than that generated at the tree level. This behavior is in agreement with previous results~\cite{Apel}, which claim that in UED models with only one extra dimension the one--loop effects dominate over the tree--level ones.

It has been pointed out~\cite{Apel} that UED models do not impact the electroweak observables at the tree level, which allows the existence of a relatively large compactification radius, that can be of order of $R^{-1}\geq 300\ \mathrm{GeV}$~\cite{Apel}. In this scenario, $|\Delta \kappa^{\rm tree}|=0.72\times 10^{-5}$ and $\Delta Q^{\rm tree}=1.44\times 10^{-5}$, whereas $|\Delta \kappa^{\rm 1-loop}|=0.47\times 10^{-4}$ and $|\Delta Q^{\rm 1-loop}|=0.73\times 10^{-4}$ for $\sqrt{Q^2}=0.5\ \mathrm{TeV}$. Since the expected sensitivity to these form factors in the ILC is of $O(10^{-4})$~\cite{LHC-ILC}, we conclude that only a compactification scale $R^{-1}\sim v$ will be at the reach of this collider.

\section{Conclusions}
\label{C}Many well-motivated standard model extensions predict the existence of new gauge bosons. Such new particles would arise by direct production if there is enough energy available, or through their virtual effects on some observables. The last scenario seems to be the most promising if the new particles have masses much larger than the Fermi scale. In this case, high precision measurements are needed in order to detect any deviation from the SM predictions. In this paper, we have examined the one-loop effects of extra dimensions on the $W^-W^+\gamma$ and $W^-W^+Z$ vertices, with the $\gamma$ and the $Z$ gauge bosons off-shell. The study of these vertices through the reaction $e^+e^-\to W^3\to W^+W^-$, will be the subject of important experimental attention at the future ILC. Our study was focused on the one-loop effects of the excited KK gauge modes that arise after compactification of a five dimensional $SU_L(2)$ Yang-Mills theory. Some details about the compactification scheme were discussed. Especial emphasis was put on the fact that the four dimensional theory satisfy two types of gauge transformations, namely, the standard gauge transformations of $SU_L(2)$ and a more complicated type of nonstandard gauge transformations. While the former is the usual gauge transformation to which are subject the zero KK modes of the four dimensional theory, the latter determines in a nontrivial way the gauge nature of the excited KK modes. A gauge-fixing procedure for the nonstandard gauge transformations that is covariant under the standard gauge transformations was introduced and the corresponding ghost sector derived. The essential pieces of the quantized excited sector of the four dimensional theory that determine the one-loop effects on light Green functions were presented. The corresponding Feynman rules were presented and used to calculate the Green function associated with the off-shell $W^-W^+W^3$ vertex. Motivated by the fact that the Pinch Technique predictions coincide, at any order of perturbation theory, with those of the Background Field Method for the especial value $\xi_Q=1$ of the gauge parameter, we have presented our results using our gauge-fixing procedure in the Feynman-'t Hooft gauge. It was found that the one-loop form factors associated with the $W^-W^+W^3$ vertex are free of ultraviolet divergences and well-behaved at high energies. In particular, it is shown that the new physics effects decouple in the limit of a small compactification scale $R$. Our numerical analysis for the $\Delta \kappa$ and $\Delta Q$ form factors shows that for energies as those planed in the ILC and a new physics scale of $R^{-1}\sim 1$ TeV, the contribution of the excited KK gauge modes to these observables are suppressed with respect to the SM radiative correction by up to two orders of magnitude in the case of $\Delta \kappa$, but it is  of the same order of magnitude in the case of $\Delta Q$. However, the one--loop KK contributions are of the same order of magnitude than those generated by new gauge bosons in other contexts of physics beyond the SM. The contribution of an operator of higher canonical dimension to these form factors also was studied. It was found that this contribution is one order of magnitude lower than that induced at the one-loop level by the KK modes, which is in agreement with previous results that point out this type of behavior in universal extra dimensional models with only one extra dimension. In general terms, for a compactification scale in the range $0.5 \ \mathrm{TeV}<R^{-1}<1.0\ \mathrm{TeV}$, the tree--level contribution to the $\Delta \kappa$ and $\Delta Q$ form factors is of about $10^{-6}$, whereas the one-loop effect of the KK modes is of order of $10^{-5}$. These values are out of the sensitivity of the ILC, which expected to bo of $O(10^{-4})$. However, if the compactification scale is of the order of the Fermi scale ($\sim 300 \ \mathrm{GeV}$), effects of one extra dimension may be at the reach of this collider.

\acknowledgments{ We acknowledge financial support from CONACYT and SNI (M\' exico). J. J. T. also acknowledges support from VIEP-BUAP under grant TOCJ-EXC10-I.}

\end{document}